\documentclass[12pt]{article}
\usepackage[english]{babel}
\usepackage{graphicx,subfigure}
\usepackage{amssymb,amsthm,amsmath,amsfonts}
\usepackage{float}
\usepackage{cite}
 
\newtheorem{remark}{Remark}
\begin{document}

\title{Operatorial formulation of crimo-taxis\\ phenomena in a street}

\author{M.~Gorgone, G.~Inferrera and C.F.~Munaf\`o\\
\ \\
{\footnotesize Department of Mathematical and Computer Sciences,}\\
{\footnotesize Physical Sciences and Earth Sciences, University of Messina}\\
{\footnotesize Viale F. Stagno d'Alcontres 31, 98166 Messina, Italy}\\
{\footnotesize mgorgone@unime.it; guinferrera@unime.it; carmunafo@unime.it}
}

\date{Submitted.}

\maketitle
	
\begin{abstract}	
In this paper, the ladder operators and a quantum-like approach are used to construct an operatorial version of a model dubbed crimo-taxis. In a classical framework, the crimo-taxis model is represented by reaction-diffusion partial differential equations describing a population divided into three interacting subgroups (ordinary citizens, drug users/dealers, and law enforcement personnel).
In this new framework, the agents of the subgroups are modeled using annihilation, creation, and number fermionic operators, and their time evolution is assumed to be ruled by a Hermitian time-independent Hamiltonian operator suitable to capture the interactions among the subgroups. 
Furthermore, a recent extension of the standard Heisenberg view, namely $(\mathcal{H},\rho)$--induced dynamics, is also taken into account. Two scenarios, characterized by different initial spatial distributions of the subgroups, are considered. The results of some numerical simulations in a one--dimensional setting are presented and discussed.
\end{abstract}



\section{Introduction}\label{sec:intro}
The collective dynamics like the one where interacting species or populations occupy a given spatial region is a crucial topic in theoretical biology, ecology, social science, \ldots \cite{epstein1997nonlinear,murray2003mathematical}. 

In this paper, we focus on a special problem connected to urban crime, where different subgroups of a population are interacting. In particular, our aim is to model the resulting dynamics where different subgroups of a population, \emph{i.e.}, ordinary citizens, drug users/dealers, and law enforcement personnel occupy a spatial region like a street; as the crime phenomena do not occur uniformly in space and time, the various subgroups, due to their local interaction and diffusion processes, can spread in space at different times.

Traditionally, the dynamics of spatially distributed groups is modeled using reaction-diffusion partial differential equations \cite{epstein1997nonlinear,
murray2003mathematical}, which can exhibit complex outcomes like pattern formation \cite{inferrera2024reaction}. In particular, a model, named \emph{crimo-taxis}, proposed by Epstein in \cite{epstein1997nonlinear}, involves three subgroups living in a one--dimensional spatial region representing a \emph{street}. In \cite{inferrera2024reaction}, it has been shown that Epstein's model does not admit  asymptotically stable equilibria, preventing one to study Turing instability \cite{Turing}; consequently, two possible variants have been proposed such that Turing instability is allowed and some interesting patterns have been exhibited. 
The two models investigated in \cite{inferrera2024reaction} can be written in compact form as
\begin{equation}
\label{react-diff}
\frac{\partial \mathbf{U}}{\partial t} = \mathbf{R}_i(\mathbf{U}) + \mathcal{D}\frac{\partial^2 \mathbf{U}}{\partial x^2},\qquad i=1,2,
\end{equation}
with
\begin{equation*}
\mathbf{U} = 
\left(
\begin{array}{c}
u \\ 
v \\ 
w
\end{array}
\right), \qquad 
\mathcal{D} = 
\left(
\begin{array}{ccc}
D_{11}  & 0       & 0 \\
-D_{21} & D_{22}  & D_{23}\\
-D_{31} & -D_{32} & D_{33}
\end{array}
\right),
\end{equation*}
and $\mathbf{R}_i(\mathbf{U})$ ($i=1,2$) being one of the following reaction terms, say
\begin{equation*}
\mathbf{R}_1(\mathbf{U}) = 
\left(
\begin{array}{c}
ru \left( 1-\frac{u}{\kappa_1} \right) - \beta uv \\ 
\beta uv - \gamma vw \\ 
- b w + \xi uvw
\end{array}
\right),
\qquad
\mathbf{R}_2(\mathbf{U}) = 
\left(
\begin{array}{c}
ru\left(1-\frac{u}{\kappa_1}\right) - \beta uv \\ 
\beta uv - \gamma v w \\ 
-bw + \xi uv \left( 1-\frac{w}{\kappa_2} \right)
\end{array}
\right),
\end{equation*}
whereas $u(t,x)$, $v(t,x)$, and $w(t,x)$ denote the densities of ordinary citizens (susceptibles), drug users/dealers (infectives), and law enforcement personnel at the time $t$ in the street position $x$, respectively; $r$ represents the susceptible reproduction rate, $\beta$ the infection rate, $\gamma$ the police arrest rate, $b$ the police natural decay rate, $\xi$ the police growth rate, whereas the coefficients $D_{ij}$ are the self-diffusion and cross-diffusion terms ($i=j$ and $i\neq j$, respectively).  
All parameters involved in  \eqref{react-diff} are assumed to be constant and positive (see \cite{epstein1997nonlinear,inferrera2024reaction} for details).

The models \eqref{react-diff} admit asymptotically stable homogeneous coexistence equilibria, with the possibility of losing their stability
due to the self-diffusion and cross-diffusion terms; moreover, the patterns exhibited by these two variants show asymptotically a stable different spatial distribution
of the three subgroups \cite{inferrera2024reaction}. 

In this paper, we propose an alternative approach using the mathematical framework of quantum mechanics \cite{roman1965quantum,merzbacher1998quantum}, specifically raising, lowering operators and number representation. In particular, we implement an operatorial model where
the agents of the system are represented by annihilation, creation and number fermionic operators whose evolution is governed by a Hermitian 
time--independent Hamiltonian $\mathcal{H}$.

The operatorial approach, initially developed in quantum mechanics, has been successfully applied to several contexts outside physics (see \cite{bagarello2012quantum,bagarello2019quantum,
bagarello2006operatorial,bagarello2007stock,
bagarello2009simplified,bagarello2009quantum,
haven2013quantum,khrennikov2010ubiquitous,
busemeyer2012quantum,asano2011quantum,asano2012quantum,
bagarello2015operator,bagarello2016first,
bagarello2016improved,khrennikova2016instability,bagarello2017h,
inferrera2022operatorial}, and references therein, where the choice of using quantum-like tools to discuss classical situations has been largely motivated). 

In this framework, we consider a system made by the three subgroups of a population living in a street, that can compete both locally and nonlocally (in the same and adiacent cells, respectively), and can spread with different diffusions; we use fermionic operators, living in a Hilbert space, to represent the agents of the model. This choice is advantageous for two main reasons: technical, because the Hilbert space is finite-dimensional, ensuring bounded operators, and practical,
because, assigning an initial condition, the mean values of the number operators can represent local densities of subgroups in different cells.

We also notice that the operatorial approaches for spatially distributed competing populations, where migration effects are included,  have already been introduced \cite{bagarello2013phenomenological,gargano2017large,gargano2021population}. However, our model description includes also nonlocal competition effects (as in \cite{inferrera2022operatorial}). Furthermore, besides the classical Heisenberg dynamics, we use the extension known as $(\mathcal{H},\rho)$--induced dynamics (see \cite{bagarello2018h,bagarello2017h,di2017operatorial}, and references therein); this method allows to study the system dynamics, ruled by the Hamiltonian, superposing periodically some \emph{rules} able to modify the values assumed by some parameters involved in the Hamiltonian (without changing its structure) according to the evolution of the system itself, thus allowing the model to adjust itself during the evolution. The rule can be seen as a way to describe the change of the attitudes of the agents during the evolution, avoiding the use of an explicitly time dependent Hamiltonian operator or the consideration of other elements into the system such as a \emph{reservoir} \cite{khrennikova2014application}, with associated analytical and computational complications.
In several applications (see \cite{bagarello2017h,di2017operatorial,bgobook}), the system approaches asymptotic equilibrium states, due to the effect of the rules. However, it can be observed that this dynamics does not necessarily manifest itself in the model investigated in this paper, even if, by using some rules, a kind of irreversibility in the dynamics is introduced.

The content of the paper is structured as follows. In Section~\ref{sec:model}, the notation is fixed, and an operatorial formulation of the crimo-taxis model is provided. A time-independent self-adjoint quadratic Hamiltonian operator rules the dynamics, accounting for both local and nonlocal diffusion and competition effects. Furthermore, the analytical solution is exhibited. In Section \ref{sec:H-rho}, the extension $(\mathcal{H},\rho)$--induced dynamics is shortly reviewed, and two suitable rules acting on inertia parameters only, and on both inertia and diffusion parameters (susceptible of having a social interpretation) are introduced. In Section \ref{sec:numericalresults}, in a one--dimensional domain, two scenarios characterized by different initial spatial distributions of the three subgroups have been considered, and some numerical results, both in the case of the standard Heisenberg view and using the $(\mathcal{H},\rho)$--induced dynamics approach, are presented. Finally, Section \ref{sec: conclusions} summarizes our conclusions, as well as further developments of the operatorial model here investigated.

\section{Operatorial framework}
\label{sec:model}
In this Section, we set the notation and construct the operatorial formulation of the crimo-taxis model, starting from a general description of its components. We introduce the fermionic operators $a_{j,\alpha}$ and $a_{j,\alpha}^\dagger$ ($j=1,2,3$, $\alpha=1,\ldots,N$) to represent the agents of the system (ordinary citizens, drug users/dealers and law enforcement personnel) in the generic cell $\alpha$ of a one--dimensional domain $\Omega$ representing a street of length $L$.
Choosing a one--dimensional setting may seem excessively restrictive, or just only an expedient to reduce the amount of computation; however, this assumption is motivated by the fact that, in natural as well social phenomena, diffusion phenomena can be often seen along well defined directions. As a consequence, considering as a spatial domain a one--dimensional street, where the three subgroups of a population are distributed, can be realistic.
The three subgroups are denoted with $\mathcal{P}_j$ ($j=1,2,3$): $\mathcal{P}_1$ refers to ordinary citizens, $\mathcal{P}_2$ to drug users/dealers, and $\mathcal{P}_3$ to law enforcement personnel. We note also that the symbol ${}^\dagger$ indicates the adjoint operation, whereas $a_{j,\alpha}$ and $a^\dagger_{j,\alpha}$ are the annihilation and creation operators, respectively. Furthermore, we introduce the self-adjoint number operators (the observables of our system) \cite{bagarello2012quantum}
\begin{equation}
\widehat{n}_{j,\alpha} = a_{j,\alpha}^\dagger a_{j,\alpha}, \qquad j=1,2,3,\qquad \alpha=1,\ldots,N.
\end{equation}			

For fermionic annihilation and
creation operators, the Canonical Anticommutation Relations (CAR) hold true, \emph{i.e.},
\begin{equation}
\label{CAR}
\{ a_{i,\alpha}, a_{j,\beta} \} = 0,\qquad \{ a^\dagger_{i,\alpha}, a^\dagger_{j,\beta} \} = 0,\qquad \{ a_{i,\alpha}, a^\dagger_{j,\beta} \} = \delta_{i,j}\,\delta_{\alpha,\beta}\, \mathcal{I},
\end{equation}
where $i,j=1,2,3$, $\;\alpha,\beta=1,\ldots,N$,  $\;\{ A,B \}:= AB + BA$ denotes the anticommutator between the operators $A$ and $B$, $\mathcal{I}$ is the identity operator, and $\delta_{i,j}$ represents the Kronecker symbol.

The ladder operators live in a Hilbert space $\mathbb{H}$ that is represented by the linear span of the orthonormal family of vectors
\begin{equation}
\label{vectors}
\varphi_{\mathbf{n}_1,\mathbf{n}_2,\mathbf{n}_3} = 
\left(\prod_{\alpha=1}^N(a_{1,\alpha}^\dagger)^{n_{1,\alpha}}\right)\left(\prod_{\alpha=1}^N(a_{2,\alpha}^\dagger)^{n_{2,\alpha}}\right)\left(\prod_{\alpha=1}^N(a_{3,\alpha}^\dagger)^{n_{3,\alpha}}\right)\varphi_{\mathbf{0},\mathbf{0},\mathbf{0}}, 
\end{equation}
where
\begin{equation}
\begin{aligned}
&\varphi_{\mathbf{n}_1,\mathbf{n}_2,\mathbf{n}_3}  \equiv \varphi_{n_{1,1},\ldots,n_{1,N},n_{2,1}\ldots,n_{2,N},n_{3,1}\ldots,n_{3,N}}, \\
&\varphi_{\mathbf{0},\mathbf{0},\mathbf{0}} \equiv \varphi_{0,\ldots,0},
\end{aligned}
\end{equation}
\emph{i.e.}, the vector $\varphi_{\mathbf{n}_1,\mathbf{n}_2,\mathbf{n}_3}$ is obtained by acting on the vacuum $\varphi_{\mathbf{0},\mathbf{0},\mathbf{0}}$ with the powers of the operators $a_{j,\alpha}^\dagger$, being $n_{j,\alpha}=0,1$ for $\alpha=1,\ldots,N$. 
Then, we get $\hbox{dim}(\mathbb{H})=2^{3N}$.  

The vector $\varphi_{\mathbf{n}_1,\mathbf{n}_2,\mathbf{n}_3}$ means that it is initially assigned a mean value equal to $n_{j,\alpha}$ to the subgroup $\mathcal{P}_j$ such that
\begin{equation}
\label{MM22} 
\widehat{n}_{j,\alpha} \varphi_{\mathbf{n}_1,\mathbf{n}_2,\mathbf{n}_3} = n_{j,\alpha} \varphi_{\mathbf{n}_1,\mathbf{n}_2,\mathbf{n}_3},
\end{equation}
\emph{i.e.}, $\varphi_{\mathbf{n}_1,\mathbf{n}_2,\mathbf{n}_3}$ is the eigenvector of the number operator $\widehat{n}_{j,\alpha}$ associated to the eigenvalue $n_{j,\alpha}$. 

Finally, the mean values of the number operators, \emph{i.e.}, 
\begin{equation}
n_{j,\alpha} = \langle\varphi_{\mathbf{n}_1,\mathbf{n}_2,\mathbf{n}_3},\widehat{n}_{j,\alpha}\varphi_{\mathbf{n}_1,\mathbf{n}_2,\mathbf{n}_3}\rangle, \qquad j = 1,2,3,\qquad\alpha=1,\ldots,N
\end{equation}
can be thought, once assigned an initial condition, as the local density of the subgroup $\mathcal{P}_j$.

\begin{figure}[h!]
\centering
\includegraphics[width=0.6\textwidth]{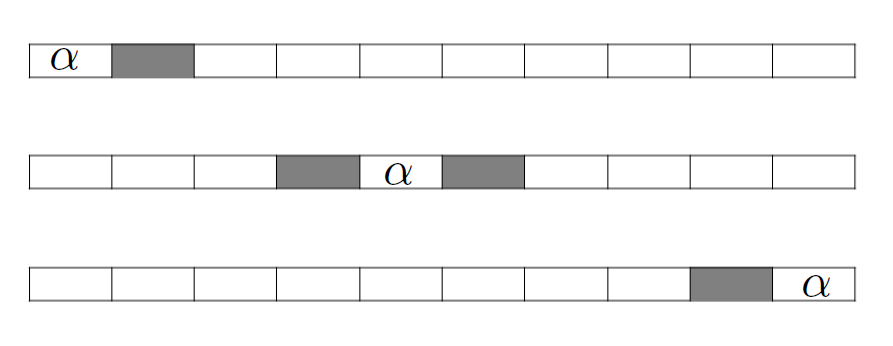}
\caption{Moore neighborhood in a one--dimensional street.}
\label{fig:Intorno} 
\end{figure}

The above operators are used to construct the Hamiltonian $\mathcal{H}$ associated to the system, which is the generator of the evolutive dynamics. In details, we assume that
\begin{equation}
\mathcal{H} = \mathcal{H}_0 + \mathcal{H}_I + \mathcal{H}_D + \mathcal{H}_C,
\end{equation}
where
\begin{equation}
\label{Hamiltonian}
\begin{aligned}
\mathcal{H}_0 &= \sum_{j=1}^{3}\sum_{\alpha=1}^{N}\omega_{j,\alpha}a_{j,\alpha}^\dagger a_{j,\alpha}, \\
\mathcal{H}_I &= \sum_{\alpha=1}^{N}\left(\lambda_{1,\alpha}(a_{1,\alpha} a_{2,\alpha}^\dagger + a_{2,\alpha} a_{1,\alpha}^\dagger)+ \lambda_{2,\alpha}(a_{2,\alpha} a_{3,\alpha}^\dagger + a_{3,\alpha} a_{2,\alpha}^\dagger)\right),\\
\mathcal{H}_D &= \sum_{j=1}^{3}\sum_{\alpha=1}^N\left(\mu_{j,\alpha}\sum_{\beta=1}^N p_{\alpha,\beta}(a_{j,\alpha} a_{j,\beta}^\dagger+a_{j,\beta} a_{j,\alpha}^\dagger)\right),\\
\mathcal{H}_C &= \sum_{\alpha,\beta=1}^{N}p_{\alpha,\beta}\left(\nu_{1,\alpha}(a_{1,\alpha} a_{2,\beta}^\dagger+a_{2,\beta} a_{1,\alpha}^\dagger)+\nu_{2,\alpha}(a_{2,\alpha} a_{3,\beta}^\dagger+a_{3,\beta} a_{2,\alpha}^\dagger)\right. \\
& +\left. \nu_{3,\alpha}(a_{1,\alpha} a_{3,\beta}^\dagger+a_{3,\beta} a_{1,\alpha}^\dagger)\right)
\end{aligned}
\end{equation}
are the different contributions in $\mathcal{H}$, whose description will be given below.
In the expressions \eqref{Hamiltonian}, the parameters $\omega_{j,\alpha}$, $\lambda_{j,\alpha}$, $\mu_{j,\alpha}$ and 
$\nu_{j,\alpha}$ ($j=1,\ldots,3$, $\alpha=1,\ldots,N$) are costant and positive, whereas the parameters $p_{\alpha,\beta}$ ($\alpha,\beta=1,\ldots,N$), being symmetric with respect to their indices, are expressed by
\begin{equation*}
p_{\alpha,\beta} = 
\begin{cases}
1 \quad \mbox{if } \beta\; \hbox{is adjacent to}\; \alpha, \\ 
0 \quad \hbox{elsewhere},
\end{cases}
\end{equation*}
\emph{i.e.}, the terms $p_{\alpha,\beta}$ are non-zero only in a Moore neighborhood. In Figure~\ref{fig:Intorno}, the one--dimensional Moore neighborhood is shown: the cells at the extremes, \emph{i.e.}, $\alpha= 1,N$, are adjacent to only one cell (the cells $\alpha=2$ and $\alpha=N-1$, respectively), and the cells $\alpha=2,\ldots,N-1$ have two adjacent cells (denoted with $\alpha-1$ and $\alpha+1$).

Below, we detail the quantities \eqref{Hamiltonian} entering the Hamiltonian $\mathcal{H}$:
\begin{itemize}
\item the term $\mathcal{H}_0$ is the free part of $\mathcal{H}$, and the parameters $\omega_{i,\alpha}$ therein involved are somehow related to the inertia of the corresponding agent \cite{bagarello2012quantum}; higher values of $\omega_{i,\alpha}$ correspond to a small propensity of the associated agent to not change in time, and vice versa;
\item  the term $\mathcal{H}_I$ rules the competitive interactions, occurring in the same cell, among the three subgroups. As there is no local competitive interaction between the agents of the subgroups $\mathcal{P}_1$ and $\mathcal{P}_3$, we use the parameters $\lambda_{1,\alpha}$ 
($\lambda_{2,\alpha}$, respectively) to measure the interaction rates between the operators $a_{1,\alpha}$ and $a_{2,\alpha}$ ($a_{2,\alpha}$ and $a_{3,\alpha}$, respectively). Moreover, also the term $a_{1,\alpha}a_{2,\alpha}^\dagger$ ($a_{2,\alpha}a_{3,\alpha}^\dagger$, respectively) can be interpreted as a competitive one since a \emph{particle} associated to $a_{1,\alpha}$ ($a_{2,\alpha}$, respectively) is \emph{destroyed} and a \emph{particle} associated to $a_{2,\alpha}$ ($a_{3,\alpha}$, respectively) is \emph{created}; the adjoint terms $a_{2,\alpha}a_{1,\alpha}^\dagger$ and $a_{3,\alpha}a_{2,\alpha}^\dagger$ swap the roles of the two agents; 
\item the term $\mathcal{H}_D$ accounts for the diffusion of the three subgroups in the spatial domain; the parameters $\mu_{j,\alpha}$ represent the diffusion coefficients of the $j$th subgroup in the cell $\alpha$. Also, for each agent, the terms $a_{j,\alpha}a_{j,\beta}^\dagger$ mean that a \emph{particle} associated to $a_{j,\alpha}$ is \emph{destroyed} and a \emph{particle} associated to $a_{j,\beta}$ is \emph{created} (in other words, the particle is annihilated in the cell $\alpha$ and created in the cell $\beta$, mimicking the diffusion in a neighboring cell); moreover, the adjoint terms $a_{j,\beta}a_{j,\alpha}^\dagger$ exchange the agents role;
\item the term $\mathcal{H}_C$ represents the nonlocal competitive interactions among the three subgroups, taking into account that their competition takes place only in adjacent cells; the values assumed by the parameters $\nu_{j,\alpha}$ allow us to measure the strength of these interactions. 
\end{itemize}
Thence, according to the Heisenberg view, the time evolution of the annihilation operators $a_{j,\alpha}$ is ruled by the following ordinary differential equations:
\begin{equation}
\label{sys_ode}
\dot{a}_{j,\alpha} = \textrm{i}[\mathcal{H},a_{j,\alpha}],\qquad j=1,2,3,
\end{equation}
where $[A,B]=AB-BA$ denotes the commutator between the operators $A$ and $B$. Since the Hamiltonian operator $\mathcal{H}$ is quadratic, it follows that the ordinary differential equations \eqref{sys_ode} turn out to be linear:
\begin{equation}\label{sys_ode_linear}
\begin{aligned}
\dot{a}_{1,\alpha}(t)& = \textrm{i} \left( - \omega_{1,\alpha} a_{1,\alpha} + \lambda_{1,\alpha} a_{2,\alpha} \phantom{\sum_{\beta=1}^{N}}\right.\\
&\left.+ \sum_{\beta=1}^{N}p_{\alpha\beta}((\mu_{1,\alpha} + \mu_{1,\beta})a_{1,\beta} + \nu_{1,\alpha}a_{2,\beta} + \nu_{3,\alpha}a_{3,\beta}) \right), \\
\dot{a}_{2,\alpha}(t) &= \textrm{i} \left( - \omega_{2,\alpha}a_{2,\alpha} + \lambda_{1,\alpha}a_{1,\alpha} + \lambda_{2,\alpha} a_{3,\alpha} \phantom{\sum_{\beta=1}^{N}}\right.\\
&\left.+\sum_{\beta=1}^{N} p_{\alpha\beta}\left((\mu_{2,\alpha} + \mu_{2,\beta})a_{2,\beta} + \nu_{1,\beta}a_{1,\beta}+\nu_{2,\alpha}a_{3,\beta} \right) \right), \\
\dot{a}_{3,\alpha}(t) &= \textrm{i} \left( - \omega_{3,\alpha}a_{3,\alpha} + \lambda_{2,\alpha} a_{2,\alpha} \phantom{\sum_{\beta=1}^{N}}\right.\\
&\left.+ \sum_{\beta=1}^{N}p_{\alpha\beta}((\mu_{3,\alpha} + \mu_{3,\beta})a_{3,\beta} + \nu_{2,\beta}a_{2,\beta} + \nu_{3,\beta}a_{1,\beta}) \right).
\end{aligned}
\end{equation}	
We note also that $\mathcal{H}_I$ and $\mathcal{H}_C$ describe the effects of competition and cross-diffusion typically found in classical reaction-diffusion models \cite{epstein1997nonlinear}.
In the following Remark, we underline a possible  analogy between reaction--diffusion equations and operatorial models. 
\begin{remark}
Under a suitable substitution it is possible to rewrite the system \eqref{sys_ode_linear} in the following form
\begin{equation}
\label{operatorial_crimo_taxis_sys}
\begin{aligned}
\dot a_{1,\alpha}(t)& = \textrm{i} \left( -\widetilde{\omega}_{1,\alpha}a_{1,\alpha}+2\nu_3 a_{3,\alpha}+\widetilde{\lambda}_{1,\alpha} a_{2,\alpha} + \widetilde{\mu}_1\frac{a_{1,\alpha-1}-2a_{1,\alpha}+a_{1,\alpha+1}}{\Delta x^2} \right. \\
& +\left.  \widetilde{\nu_1}\frac{a_{2,\alpha-1}-2a_{2,\alpha}+a_{2,\alpha+1}}{\Delta x^2} + \widetilde{\nu_3}\frac{a_{3,\alpha-1}-2a_{3,\alpha}+a_{3,\alpha+1}}{\Delta x^2} \right), \\		
\dot a_{2,\alpha}(t)& = \textrm{i} \left(- \widetilde{\omega}_{2,\alpha}a_{2,\alpha} + \widetilde{\lambda}_{1,\alpha} a_{1,\alpha} +\widetilde{\lambda}_{2,\alpha} a_{3,\alpha} + \widetilde{\mu}_2\frac{a_{2,\alpha-1}-2a_{2,\alpha}+a_{2,\alpha+1}}{\Delta x^2}\right.  \\
&+\left. \widetilde{\nu_1}\frac{a_{1,\alpha-1}-2a_{1,\alpha}+a_{1,\alpha+1}}{\Delta x^2} + \widetilde{\nu_2}\frac{a_{3,\alpha-1}-2a_{3,\alpha}+a_{3,\alpha+1}}{\Delta x^2} \right), \\
\dot a_{3,\alpha}(t) &= \textrm{i} \left( -\widetilde{\omega}_{3,\alpha}a_{3,\alpha}+2\nu_3 a_{1,\alpha} + \widetilde{\lambda}_{2,\alpha} a_{2,\alpha} + \widetilde{\mu}_3\frac{a_{3,\alpha-1}-2a_{3,\alpha}+a_{3,\alpha+1}}{\Delta x^2}\right.\\
& +\left. \widetilde{\nu_2}\frac{a_{2,\alpha-1}-2a_{2,\alpha}+a_{2,\alpha+1})}{\Delta x^2} + \widetilde{\nu_3}\frac{a_{1,\alpha-1}-2a_{1,\alpha}+a_{1,\alpha+1}}{\Delta x^2} \right),
\end{aligned}
\end{equation}
where
\begin{equation}
\begin{aligned}
&\widetilde{\omega}_{i} = \omega_{i}-4\mu_i, &&\qquad
\widetilde{\lambda}_{i,\alpha} = \lambda_{\alpha} + 2\nu_i, \\
&\widetilde{\mu}_i = 2\mu_i \Delta x^2, &&\qquad
\widetilde{\nu_i} = \nu_i \Delta x^2.
\end{aligned}
\end{equation}
Particular emphasis is placed on the last three terms of each equation in the system \eqref{operatorial_crimo_taxis_sys}, which can be interpreted as the spatial discretizations of the second-order spatial derivative (with step size $\Delta x$). These terms represent the self-diffusion and cross-diffusion, respectively. Therefore, an operatorial model of a lattice can be thought of as the spatial discretization of a system of partial differential equations of  reaction-diffusion type.
\end{remark}

\subsection{Analytical solution}
In this Subsection, we provide an explicit form for the unknowns involved in the system \eqref{sys_ode_linear}; these are operators expressed by $2^{3N}\times 2^{3N}$ matrices, thus implying that, without any further simplification, $3N\cdot 2^{6N}$ scalar ordinary differential equations have to be solved in a complex domain. However, since the system \eqref{sys_ode_linear} is made of linear differential equations, the computational complexity can be drastically reduced. Let us collect the $3N$ operators $a_{j,\alpha}$ in the column vector $\mathbf{A}$, say
\begin{equation*}
\mathbf{A} = \left[ a_{1,1},\ldots,a_{1,N},a_{2,1},\ldots,a_{2,N},a_{3,1},\ldots,a_{3,N} \right]^T,
\end{equation*}
(where $^T$ denotes transposition), and consider a $3N\times 3N$ matrix $\Gamma$ (whose real entries, after the number of agents $N$ is fixed, can be derived from equations \eqref{sys_ode_linear}); then, the ordinary differential equations ruling the evolution of the annihilation operators can be read in the following compact form
\begin{equation}
\label{compact_annihilation}
\frac{d\mathbf{A}}{dt} = \textrm{i} \Gamma\, \mathbf{A},
\end{equation}
whose analytical solution is expressed by
\begin{equation}\label{solution_compact}
\mathbf{A}(t)=\exp\left(\textrm{i}\,\Gamma t\right)\mathbf{A}_0=
B(t)\mathbf{A}_0.
\end{equation}
Here, we denote with $n_{j,\alpha}^0$ the $j$th subgroup density in the cell $\alpha$ at the initial time $t_0=0$,
and with $\mathbf{n}^0$ the corresponding initial value, say
\begin{equation*}
\mathbf{n}^0 = \left({n_{1}^0},\ldots,{n_{3N}^0}\right)^T,
\end{equation*}
where $n_k^0$ is the initial density of $j$th subgroup at the cell $\alpha$,  with $\allowbreak k=(j-1)N+\alpha$.

Given the generic entry $B_{j,k}$ of the $3N\times 3N$ matrix $B(t)$, using the relation 
\[
n_{j,\alpha}(t) =\langle\varphi_{\mathbf{n}_1,\mathbf{n}_2,\mathbf{n}_3},a^\dagger_{j,\alpha}(t)a_{j,\alpha}(t)\varphi_{\mathbf{n}_1,\mathbf{n}_2,\mathbf{n}_3}\rangle, \qquad j = 1,2,3,\qquad\alpha=1,\ldots,N,
\]
and the CAR \eqref{CAR}, the mean values of the number operators at time $t$ can be easily recovered \cite{inferrera2022operatorial}:
\begin{equation}
n_{j,\alpha}(t) = \sum_{\beta=1}^{3N}\left|B_{(j-1)N+\alpha,\beta}(t)\right|^2n_{\beta}^0,\qquad j = 1,2,3,\qquad\alpha=1,\ldots,N.
\end{equation}
The approach above described reveals really useful but has some limitations, as detailed in the following Remark.
\begin{remark}\
\begin{itemize}
\item A first integral is possessed by the Hamiltonian $\mathcal{H}$, say
\begin{equation*}
\left[\mathcal{H},\sum_{\alpha=1}^N\left(\widehat{n}_{1,\alpha} +\widehat{n}_{2,\alpha}+\widehat{n}_{3,\alpha}\right)\right]=0,
\end{equation*}
\emph{i.e.}, sum of the three subgroup densities all over the cells of the entire domain remains constant in time.
\item The Hamiltonian has a quadratic structure: as a consequence, the dynamics in each cell is at most quasiperiodic.
\end{itemize}
\end{remark}
An alternative strategy to obtain different (especially not quasiperiodic) dynamical behaviors could be that of incorporating higher order terms in the Hamiltonian, ora considering an open quantum system \cite{bagarello2019quantum,bagarello2018non}. Unfortunately, the computational difficulties will increase because a very huge number of differential equations must be numerically solved (see \cite{bagarello2018h}). A recent approach, with the goal of enriching the dynamics without increasing the computational complexity, has been introduced in \cite{bagarello2018h,bagarello2017h,di2017operatorial,di2017political,
bagarello2017modeling}, where non-oscillatory trends can be recovered still keeping an Hamiltonian with a time-independent and quadratic structure. 

\section{The ($\mathcal{H},\rho$)--induced dynamics approach}
\label{sec:H-rho}
The aim of the $(\mathcal{H},\rho)$--induced dynamics approach is to avoid the quasiperiodic behavior, that is due to the classical Heisenberg view, by including in the Hamiltonian some effects: this is done using suitable \emph{rules}. More in detail, a rule is a law that changes, at fixed instants, some of the values of the parameters entering the Hamiltonian, so that the model is somehow able to adapt itself according to the time evolution of the system. From a mathematical viewpoint, a rule is a mapping $\rho\colon\mathbb{R}^p\rightarrow\mathbb{R}^p$ working on the $p$ parameters included in the Hamiltonian. Essentially, the main advantage of this mathematical expedient is to allow the agents to modify, according to their evolution, their attitudes and interaction rates.

Within the framework of $(\mathcal{H},\rho)$--induced dynamics  (see \cite{bagarello2018h} for a complete description), the values of some parameters involved in the Hamiltonian are stepwise constant in time, as periodically they are \emph{adjusted} according to the evolution of the system itself. From a computational point of view, this implies that the complete evolution of the system in the time interval $[0,T]$ is obtained by glueing the local evolutions in a finite number of subintervals. In each subinterval the Hamiltonian is time-independent, but globally we have a time-dependent Hamiltonian (see \cite{bagarello2018h} for a comparison of this approach with the general case of a time-dependent Hamiltonian).

According to this method,  we fix the length $\tau$ of the subintervals, and, then, update at each instant $k\tau$ ($k$ integer) the values of the parameters as specified below.
At first, let us introduce the quantities
\begin{equation}
\begin{aligned}
&\delta_{j,\alpha}^{(k)}=n_{j,\alpha}(k\tau)-n_{j,\alpha}((k-1)\tau), \qquad j=1,2,3,\quad \alpha=1,\ldots,N,\\
&\delta_{\alpha}^{(k)} =n_{2,\alpha}(k\tau) n_{1,\alpha}(k\tau)-n_{2,\alpha}((k-1)\tau)n_{1,\alpha}((k-1)\tau),
\end{aligned}
\end{equation}
where $\delta_{j,\alpha}^{(k)}$ denotes the variation of the mean values of the number operators in the cell $\alpha$ associated to each subgroup $\mathcal{P}_j$, \emph{i.e.}, how the subgroup local densities vary at each step, and $\delta_{\alpha}^{(k)}$ is the difference between the products of the mean values of the number operators associated to the subgroups $\mathcal{P}_1$ and $\mathcal{P}_2$, \emph{i.e.}, it represents the variation at each step of the interactions between ordinary citizens and drug addicts.

More in detail, let us consider the following rules for the three families of inertia and diffusion parameters, respectively, say
\begin{subequations}
\label{rule_inertia}
\begin{align}
&\omega_{1,\alpha} \, \rightarrow \, \omega_{1,\alpha} \left( 1-\delta_{2,\alpha}^{(k)} + \delta_{3,\alpha}^{(k)} \right), \label{rule_inertia_1} \\
&\omega_{2,\alpha} \, \rightarrow \, \omega_{2,\alpha} \left( 1+\delta_{1,\alpha}^{(k)} - \delta_{3,\alpha}^{(k)} \right), \label{rule_inertia_2} \\
&\omega_{3,\alpha} \, \rightarrow \, \omega_{3,\alpha} \left( 1-(\delta_{3,\alpha}^{(k)} + \delta_{\alpha}^{(k)}) \right),\label{rule_inertia_3}
\end{align}
\end{subequations}
and
\begin{subequations}\label{rule_diffusion}
\begin{align}
&\mu_{1,\alpha} \, \rightarrow \, \mu_{1,\alpha} \left( 1+\delta_{2,\alpha}^{(k)}-\delta_{3,\alpha}^{(k)} \right),
\label{rule_diffusion_1} \\
&\mu_{2,\alpha} \, \rightarrow \, \mu_{2,\alpha} \left( 1-\delta_{1,\alpha}^{(k)} + \delta_{3,\alpha}^{(k)} \right), \label{rule_diffusion_2} \\
&\mu_{3,\alpha} \, \rightarrow \, \mu_{3,\alpha} \left( 1-\delta_{\alpha}^{(k)} \right),
\label{rule_diffusion_3}
\end{align}
\end{subequations}	
holding for all $\alpha=1,\ldots,N$.

Some comments about the rules \eqref{rule_inertia} and \eqref{rule_diffusion} are in order.
The relation \eqref{rule_inertia_1} indicates that the inertia parameter related to the subgroup $\mathcal{P}_1$  increases (or decreases) when the difference between the local densities of the subgroups $\mathcal{P}_3$ and $\mathcal{P}_2$ in the subinterval of length $\tau$ increases (or decreases). This rule suggests that an increase in the cell $\alpha$ of this difference  induces ordinary citizens to lower their tendency to change (police local density is greater than drug addicts one, so ordinary citizens would feel safe), whereas a decrease causes ordinary citizens to be less conservative (drug addicts local density is greater than police one, so ordinary citizens would not feel safe).

The relation \eqref{rule_inertia_2} is similar to rule \eqref{rule_inertia_1}: the inertia parameter related to the subgroup $\mathcal{P}_2$ increases (or decreases) as the difference between local densities of the subgroups $\mathcal{P}_1$ and $\mathcal{P}_3$  increases (or decreases). This rule takes into account that, when the above difference increases, the drug addicts are induced to remain in the same area (ordinary citizens local density is greater than  police one, therefore drug addicts continue to commit crime), whereas a decrease causes drug addicts to be less conservative (police local density  is greater than the ordinary citizens one, so that the drug addicts will tend to move to other areas to commit crimes).

The rule for inertia parameters associated to the third subgroup $\mathcal{P}_3$, as expressed in \eqref{rule_inertia_3}, incorporates the concept of social alarm introduced by Epstein \cite{epstein1997nonlinear}. The parameters update is influenced by the sum of interactions variation between ordinary citizens and drug addicts, alongside police local density. This social alarm is crucial for illustrating scenarios where law enforcement intervention is necessary to restore order.

The relation \eqref{rule_diffusion_1} indicates that the diffusion parameter related to ordinary citizens $\mathcal{P}_1$ increases (or decreases) when the difference between drug addicts and police  local densities increases (or decreases). The idea behind it is that ordinary citizens must diffuse to neighboring territories when there is a high crime rate (the difference is positive), but if the law enforcement personnel is sufficient in number to maintain order (the difference is negative), ordinary citizens can remain in their safe place.
	
The relation \eqref{rule_diffusion_2} shows that the diffusion parameter associated to drug addicts $\mathcal{P}_2$ increases (or decreases) as the difference between police and ordinary citizens local densities increases (or decreases). Drug addicts are less likely to dissolve when ordinary citizens density is high and police one is low, as this encourages continued criminal activity. In fact, with a high police presence, drug addicts may relocate to seek opportunities to commit crimes elsewhere.
	
Finally, the relation \eqref{rule_diffusion_3} reveals that the diffusion parameter related to police $\mathcal{P}_3$ increases (or decreases) when the interactions between the drug addicts and ordinary citizens decrease (or increase). This implies that police should spread  to other cells when social alarm in the cell $\alpha$ is low.

\section{Numerical results}
\label{sec:numericalresults}

In this Section, we exhibit and discuss some numerical simulations for the operatorial model \eqref{sys_ode_linear}. In the first part, we neglect spatial inhomogeneities, \emph{i.e.}, all the interactions occur just in a single cell, whereas, the second one concerns with the model investigated in a one--dimensional spatial domain made of $N=50$ cells.
In both parts, the numerical results obtained following the standard Heisenberg view and by means of the ($\mathcal{H},\rho$)--induced dynamics are compared.

\subsection{Model in one cell}
\label{nospa}
In this case, the time evolution is investigated in the situation where all the population is confined in only one cell: as the diffusion effects and the nonlocal competitive interactions are absent, the Hamiltonian operator reduces to the form $\mathcal{H}=\mathcal{H}_0+\mathcal{H}_I$.

\begin{figure}[h!]
\centering
\subfigure[]{\includegraphics[width=0.49\textwidth]{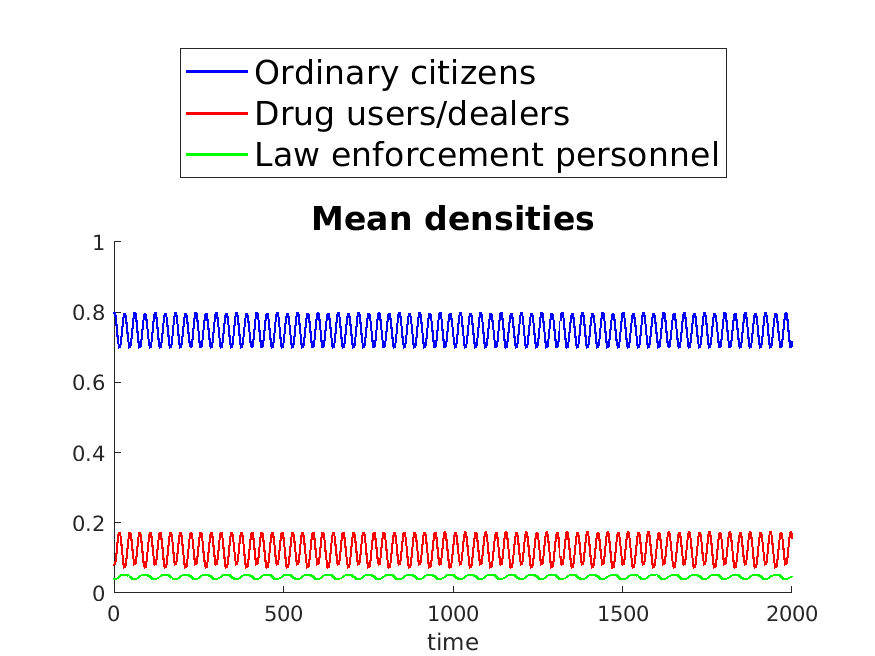}\label{fig:3modia-first}}
\subfigure[]{\includegraphics[width=0.49\textwidth]{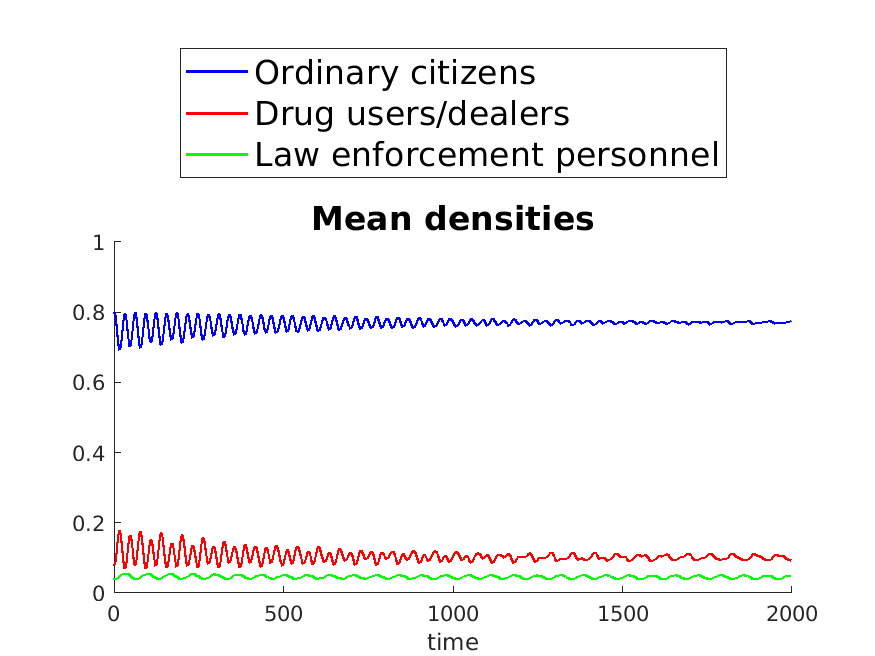}\label{fig:3modib-first}}
\caption{Time evolution, in one cell, of the mean values of the number operators related to the three subgroups. No rule (Figure~\ref{fig:3modia-first}), and  rule~\eqref{rule_inertia} with $\tau=4$ (Figure~\ref{fig:3modib-first}). The initial conditions in both cases are: $n_1^0=0.8$, $n_2^0=0.08$, $n_3^0=0.04$. The inertia and interaction parameters are $\omega_{1}=0.5$, $\omega_{2}=0.3$, $\omega_{3}=0.2$, and $\lambda_1=0.04$, $\lambda_2=0.025$, respectively.}
\label{fig:3modi}
\end{figure}

\begin{figure}[h!]
\centering
\subfigure[]{\includegraphics[width=0.49\textwidth]{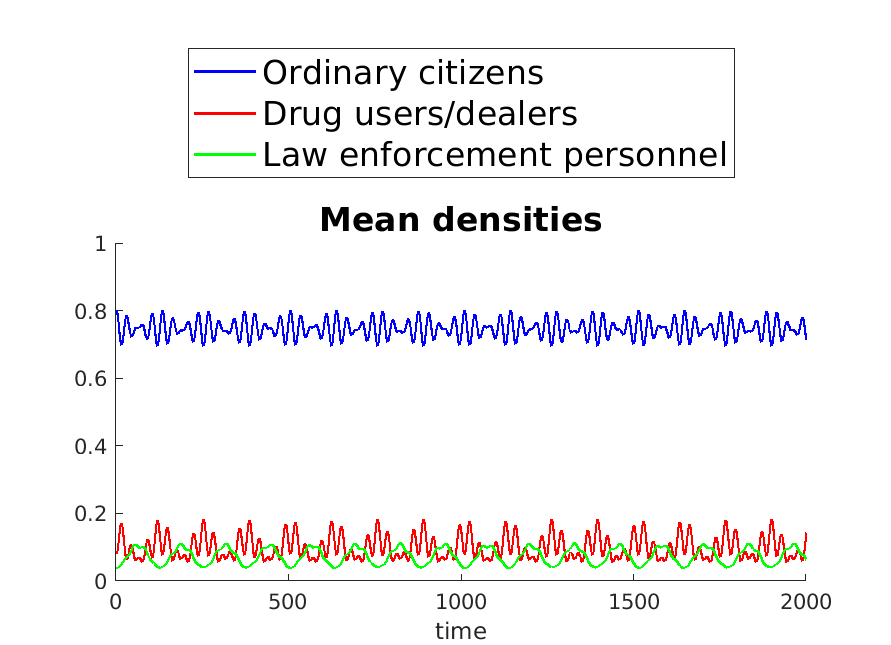}\label{fig:3modia-second}}
\subfigure[]{\includegraphics[width=0.49\textwidth]{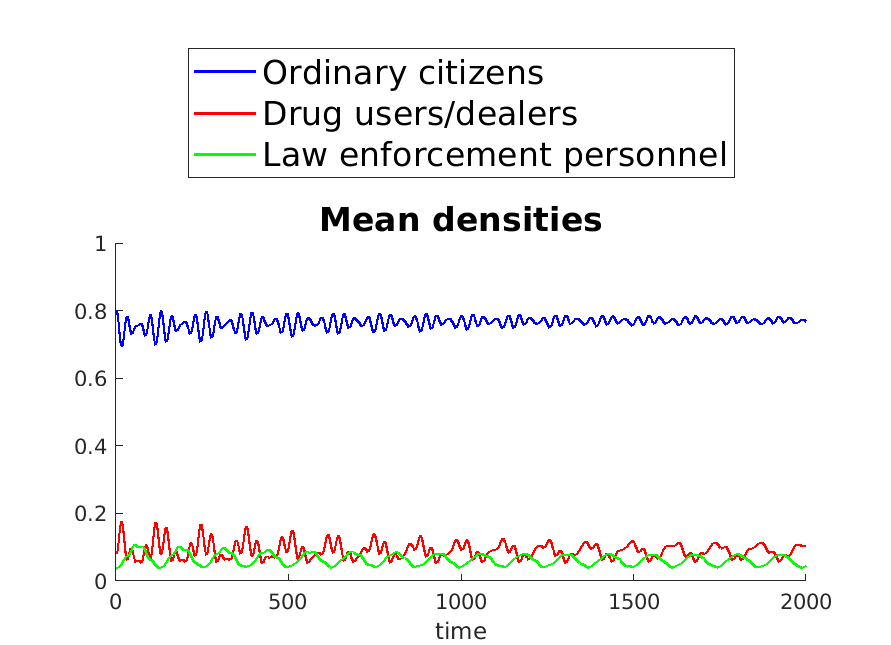}\label{fig:3modib-second}}
\caption{Time evolution, in one cell, of the mean values of the number operators related to the three subgroups. No rule (Figure~\ref{fig:3modia-second}), and  rule~\eqref{rule_inertia} with $\tau=4$ (Figure~\ref{fig:3modib-second}). The initial conditions in both cases are: $n_1^0=0.8$, $n_2^0=0.08$, $n_3^0=0.04$. The inertia and interaction parameters are $\omega_{1}=0.3$, $\omega_{2}=0.1$, $\omega_{3}=0.1$, and $\lambda_1=0.04$, $\lambda_2=0.025$, respectively.}
\label{fig:3modi-second}
\end{figure}

\begin{figure}[h!]
\centering
\subfigure[]{\includegraphics[width=0.49\textwidth]{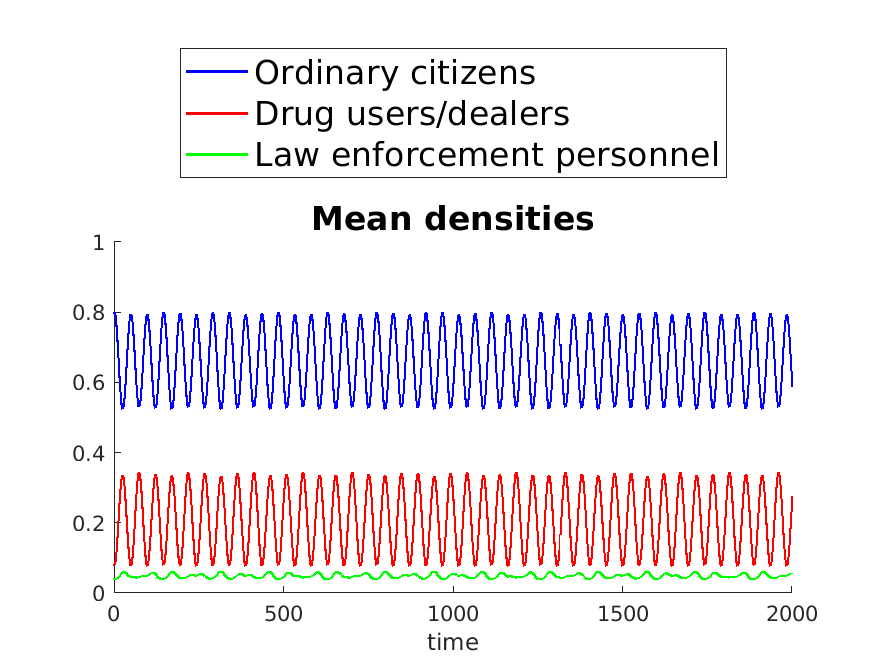}\label{fig:3modia-third}}
\subfigure[]{\includegraphics[width=0.49\textwidth]{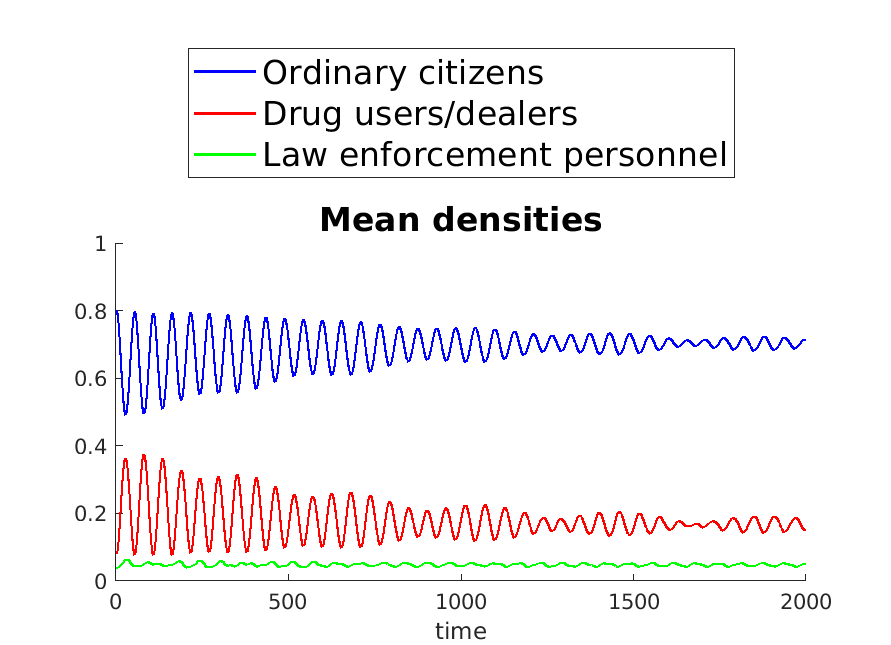}\label{fig:3modib-third}}
\caption{Time evolution, in one cell, of the mean values of the number operators related to the three subgroups. No rule (Figure~\ref{fig:3modia-third}), and  rule~\eqref{rule_inertia} with $\tau=4$ (Figure~\ref{fig:3modib-third}). The initial conditions in both cases are: $n_1^0=0.8$, $n_2^0=0.08$, $n_3^0=0.04$. The inertia and interaction parameters are $\omega_{1}=0.2$, $\omega_{2}=0.1$, $\omega_{3}=0.3$, and $\lambda_1=0.04$, $\lambda_2=0.025$, respectively.}
\label{fig:3modi-third}
\end{figure}

In the standard Heisenberg dynamics, we see in Figures~\ref{fig:3modia-first}, \ref{fig:3modia-second} and \ref{fig:3modia-third} that the evolution exhibits a never ending oscillatory trend, as expected. On the contrary, when the rule \eqref{rule_inertia} is used, the amplitude of the oscillations is decreasing in time, and the mean densities of the three subgroups tend to approach an asymptotic equilibrium (see Figures~\ref{fig:3modib-first}, \ref{fig:3modib-second} and \ref{fig:3modib-third}). Furthermore, from Figures~\ref{fig:3modib-first}, \ref{fig:3modib-second} and \ref{fig:3modib-third}, an interpretation can be the following: law enforcement personnel is able to better keep drug users/dealers under control. We remark that the time required to reach a stationary trend after the transient evolution depends on the choosen parameters.

\subsection{Spatial model}
\begin{figure}[H]
\centering
\includegraphics[width=0.49\textwidth]{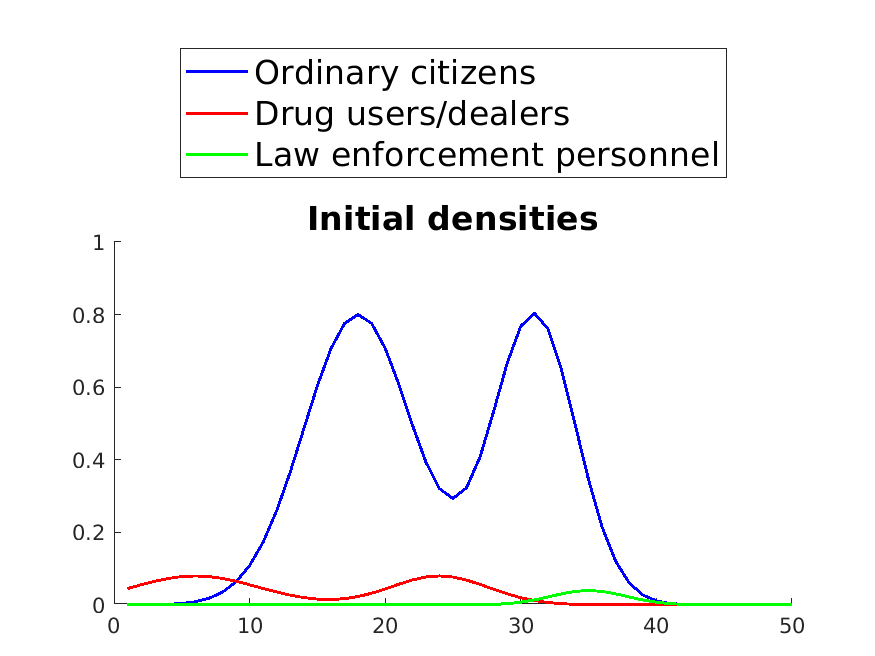}
\includegraphics[width=0.49\textwidth]{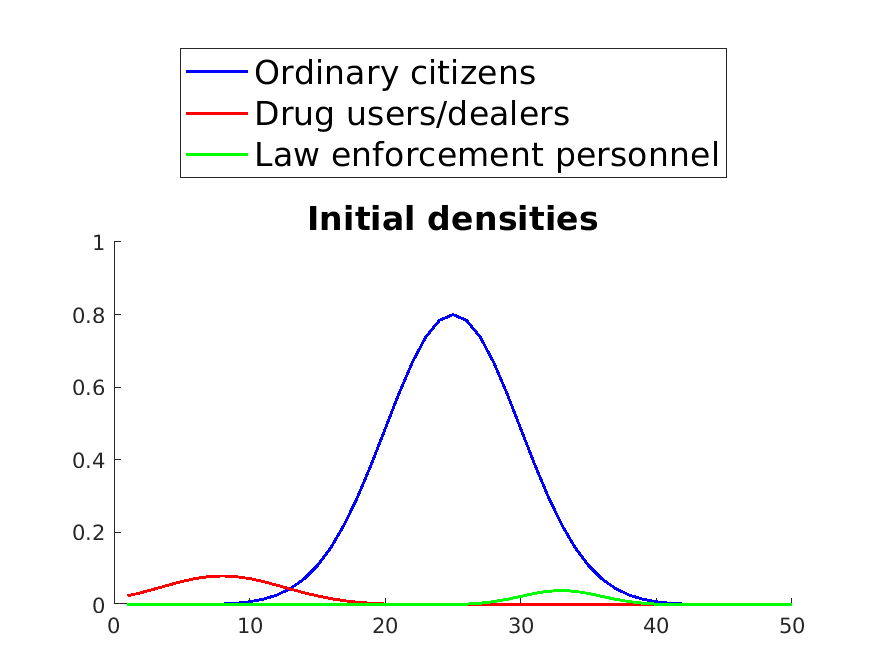}
\caption{Initial conditions in a one--dimensional street made of $N=50$ cells for the first \eqref{druginit2} and second \eqref{druginit3} scenario, respectively.}
\label{fig:cond_ini}
\end{figure}

Let us consider a street represented by a one--dimensional region $\Omega$ divided in $N=50$ adjacent cells, and investigate two scenarios characterized by different initial spatial distributions of the subgroups. The two situations are described as follows:
\begin{description}
\item[first scenario:] there are two areas with a high density of ordinary citizens, two suburbs with a small concentration of drug users/dealers, and a very small and dislocated area, with respect to the center of the street, where law enforcement personnel are concentrated. This scenario can be represented by the following initial mean densities:
\begin{equation}
\label{druginit2}
\begin{aligned}
&n_{1,\alpha}^0=0.8\left(\exp(-(\alpha-18)^2/32)+\exp(-(\alpha-31)^2/19)\right), \\
&n_{2,\alpha}^0=0.08\left(\exp(-(\alpha-6)^2/44)+\exp(-(\alpha-24)^2/26)\right),&& \alpha=1,\ldots,50,\\
&n_{3,\alpha}^0=0.04\exp(-(\alpha-35)^2/15);
\end{aligned}
\end{equation}
\item[second scenario:] there is an area with a high density of ordinary citizens, and two small suburban areas occupied by drug users/dealers and law enforcement personnel, respectively. This scenario is similar to the one considered in \cite{epstein1997nonlinear}. We choose the following initial mean densities:
\begin{equation}
\label{druginit3}
\begin{aligned}
&n_{1,\alpha}^0=0.8(\exp(-(\alpha-25)^2/50), \\
&n_{2,\alpha}^0=0.08(\exp(-(\alpha-8)^2/42), \qquad\qquad \alpha=1,\ldots,50,\\
&n_{3,\alpha}^0=0.04\exp(-(\alpha-33)^2/17).
\end{aligned}
\end{equation}
\end{description}
The initial densities corresponding to the three subgroups for both scenarios are shown in Figure~\ref{fig:cond_ini}.
Now, let us fix the initial parameters entering the model.
We choose the following two sets of inertia parameters
\begin{equation}
\label{Inertia1}
\begin{aligned}
&\omega_{1,\alpha}=0.5 + 0.25\left(\exp(-(\alpha-18)^2/32)+\exp(-(\alpha-31)^2/19)\right),\\
&\omega_{2,\alpha}=0.2+0.1\left(\exp(-(\alpha-6)^2/44)+\exp(-(\alpha-24)^2/26)\right),\\
&\omega_{3,\alpha}=0.3+0.15\exp(-(\alpha-35)^2/15),
\end{aligned}
\end{equation}
and
\begin{equation}
\label{Inertia2}
\begin{aligned}
&\omega_{1,\alpha}=0.5 + 0.25\exp\left(-(\alpha-25)^2/50)\right),\\
&\omega_{2,\alpha}=0.2 + 0.1\exp\left(-(\alpha-8)^2/42)\right),\\
&\omega_{3,\alpha}=0.3 + 0.15\exp\left(-(\alpha-33)^2/17)\right),
\end{aligned}
\end{equation}
corresponding to the first and second scenario, respectively. The choices \eqref{Inertia1} and \eqref{Inertia2} describe the fact that each agent can be thought to have an inertia similar to the initial spatial distribution of the corresponding subgroup.

Finally, we assign the interaction, diffusion and cross-diffusion parameters (as chosen in the model described in \cite{epstein1997nonlinear,inferrera2024reaction}), say 
\begin{equation}
\label{drugpar1}
\begin{array}{lll}
\lambda_{1,\alpha}=0.04, &\qquad \mu_{1,\alpha}=0.07(1+\xi_1),&\qquad\nu_{1,\alpha}=0.006,  \\
\lambda_{2,\alpha}=0.025, &\qquad \mu_{2,\alpha}=0.05(1+\xi_1), &\qquad \nu_{2,\alpha}=0.006, \\
\phantom{g}&\qquad \mu_{3,\alpha}=0.035(1+\xi_2), &\qquad \nu_{3,\alpha}=0.003,
\end{array}
\end{equation}
where $\xi_1\in[0,0.01]$ and $\xi_2\in[0,0.002]$ are small random perturbations. The constants $\xi_1$ and $\xi_2$ are introduced such that all agents may have different initial mobilities.

\begin{figure}[h!]
\centering
\subfigure[]{\includegraphics[width=0.32\textwidth]{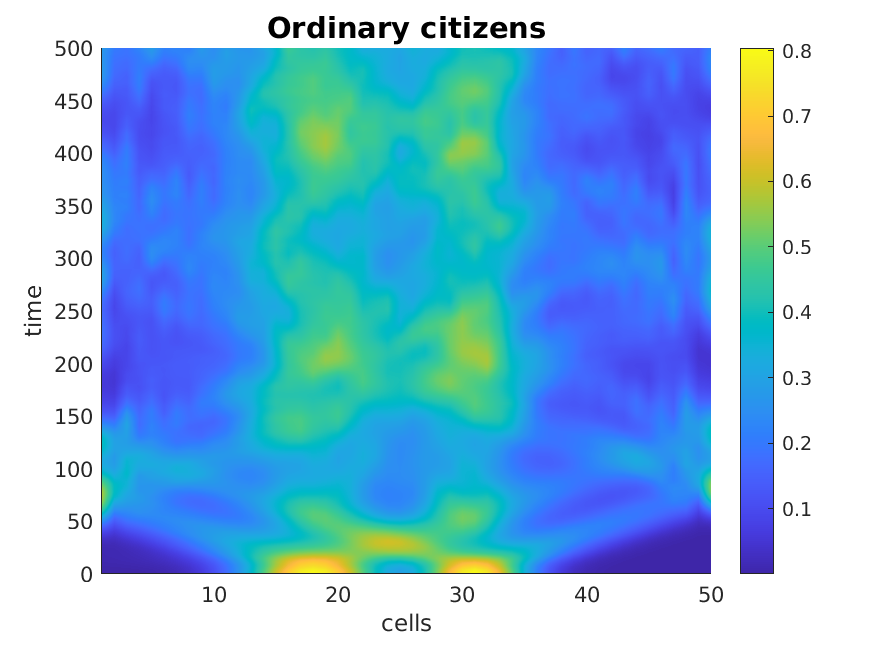}\label{fig:ic-2-om2-druganorule-a}}
\subfigure[]{\includegraphics[width=0.32\textwidth]{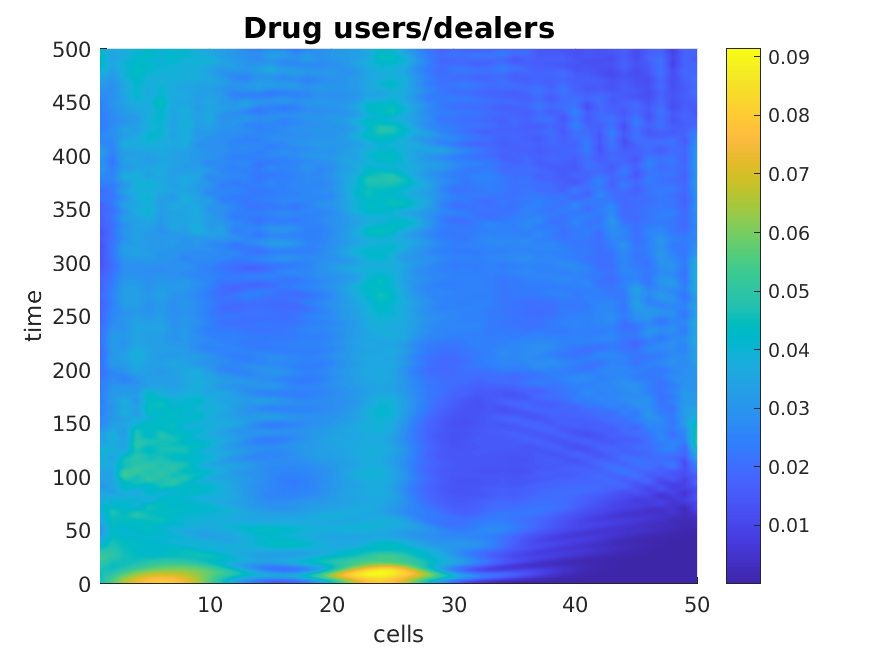}\label{fig:ic-2-om2-druganorule-b}}
\subfigure[]{\includegraphics[width=0.32\textwidth]{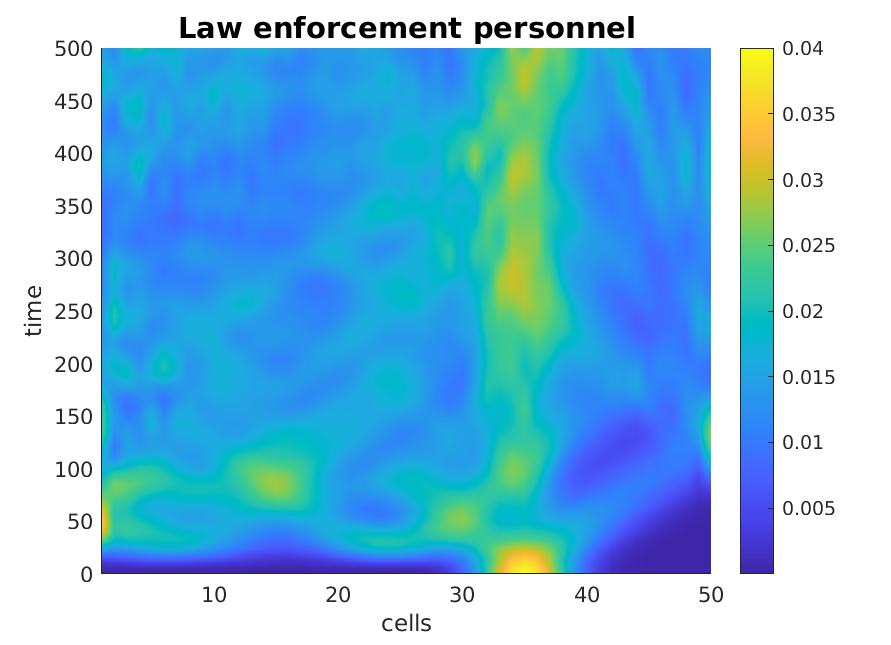}\label{fig:ic-2-om2-druganorule-c}}\\
\subfigure[]{\includegraphics[width=0.32\textwidth]{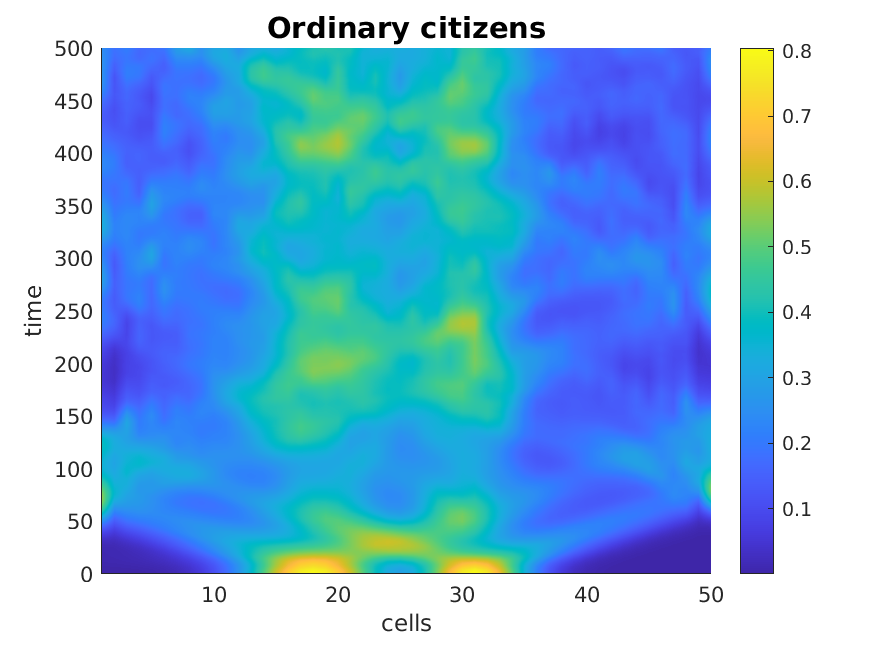}\label{fig:ic-2-om2-druganorule-d}}
\subfigure[]{\includegraphics[width=0.32\textwidth]{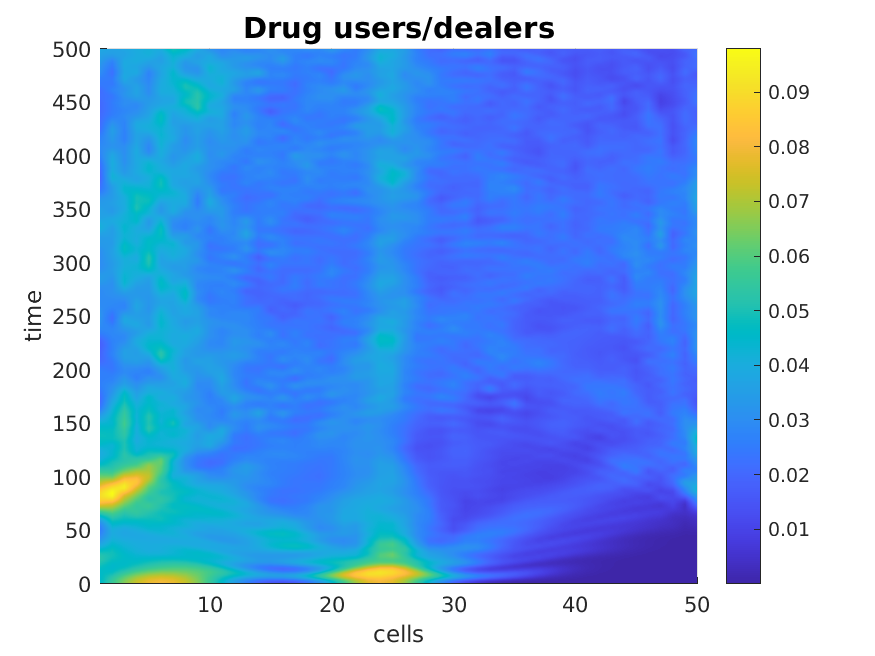}\label{fig:ic-2-om2-druganorule-e}}
\subfigure[]{\includegraphics[width=0.32\textwidth]{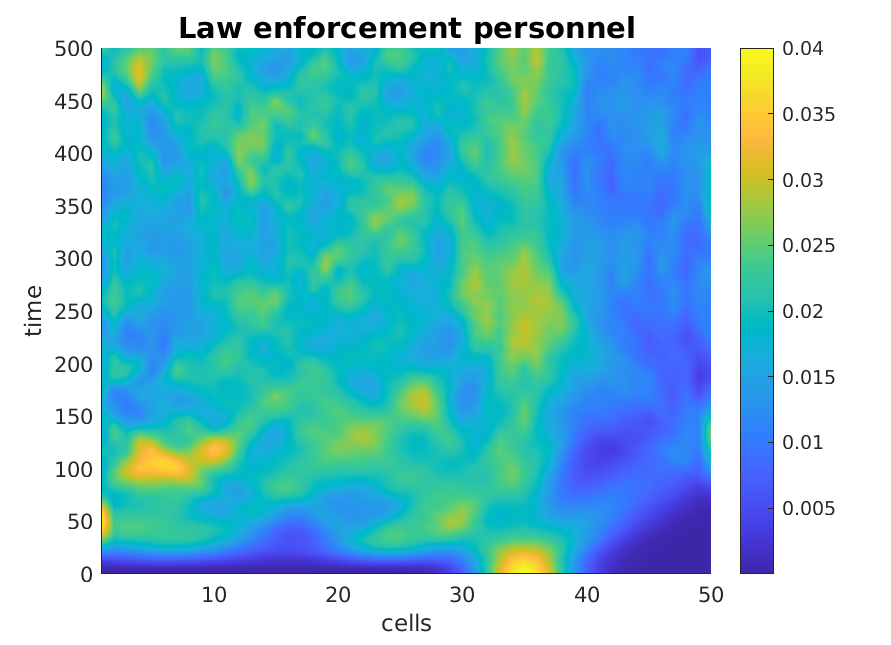}\label{fig:ic-2-om2-druganorule-f}}\\
\subfigure[]{\includegraphics[width=0.32\textwidth]{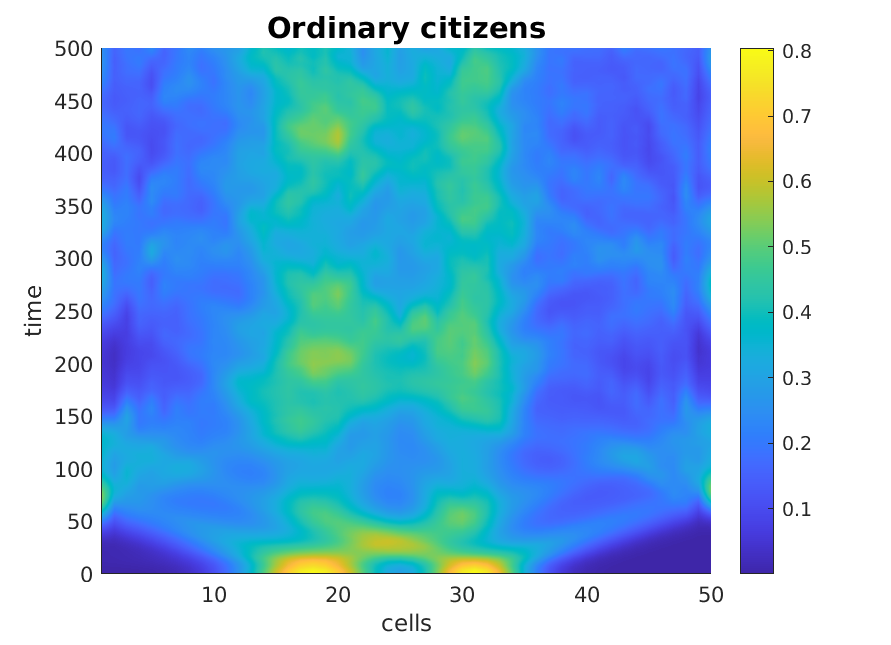}\label{fig:ic-2-om2-druganorule-g}}
\subfigure[]{\includegraphics[width=0.32\textwidth]{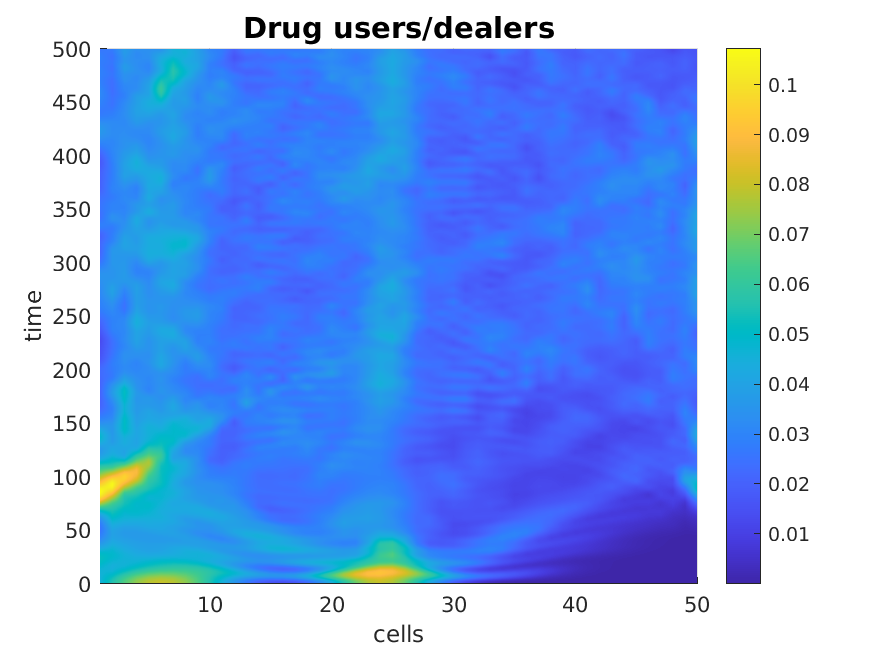}\label{fig:ic-2-om2-druganorule-h}}
\subfigure[]{\includegraphics[width=0.32\textwidth]{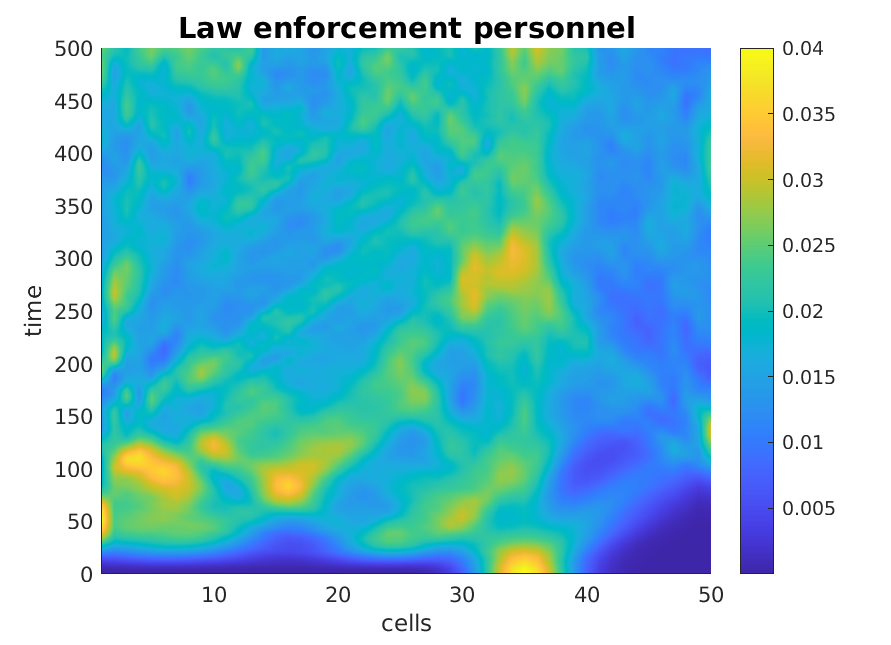}\label{fig:ic-2-om2-druganorule-i}}
\caption{First scenario: time evolution and spatial distribution of the three subgroup densities. Ordinary citizens (\ref{fig:ic-2-om2-druganorule-a}, \ref{fig:ic-2-om2-druganorule-d} and \ref{fig:ic-2-om2-druganorule-g}), drug users/dealers (\ref{fig:ic-2-om2-druganorule-b}, \ref{fig:ic-2-om2-druganorule-e} and \ref{fig:ic-2-om2-druganorule-h}), and law enforcement personnel (\ref{fig:ic-2-om2-druganorule-c}, \ref{fig:ic-2-om2-druganorule-f} and \ref{fig:ic-2-om2-druganorule-i}). No-rule (upper), rule \eqref{rule_inertia} (center) and rules \eqref{rule_inertia}--\eqref{rule_diffusion} (bottom). The values of initial mean densities and parameters are given in \eqref{druginit2}, \eqref{Inertia1} and \eqref{drugpar1}.}
\label{fig:ic-2-om2-druganorule}
\end{figure}

\begin{figure}[h!]
\centering
\subfigure[]{\includegraphics[width=0.32\textwidth]{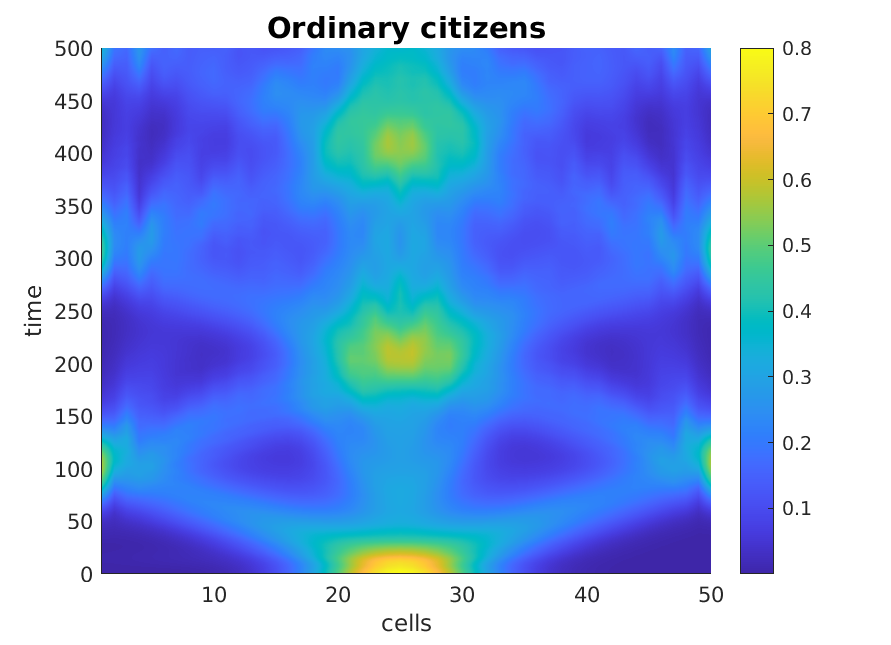}\label{fig:ic-3-om2-druganorule-a}}
\subfigure[]{\includegraphics[width=0.32\textwidth]{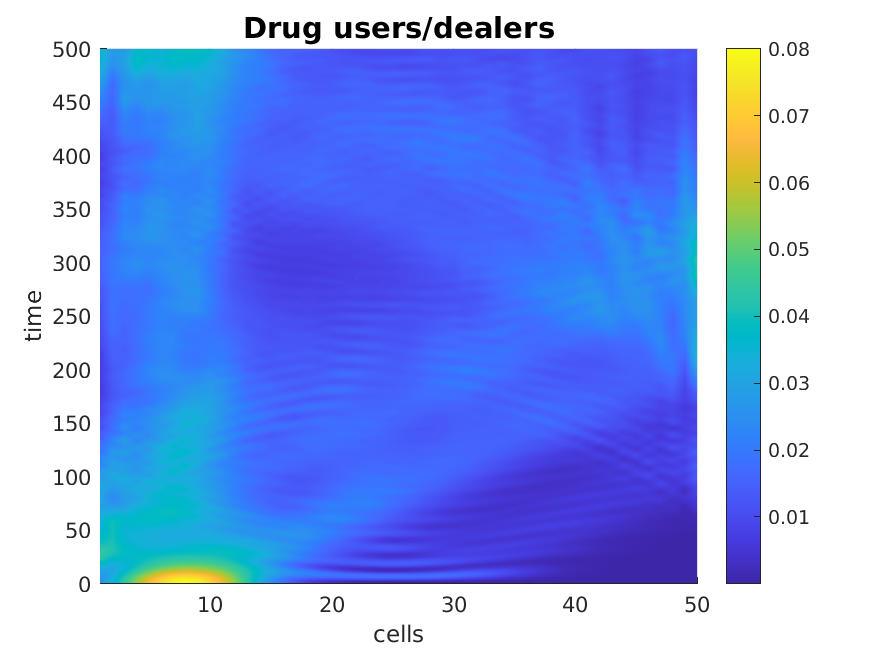}\label{fig:ic-3-om2-druganorule-b}}
\subfigure[]{\includegraphics[width=0.32\textwidth]{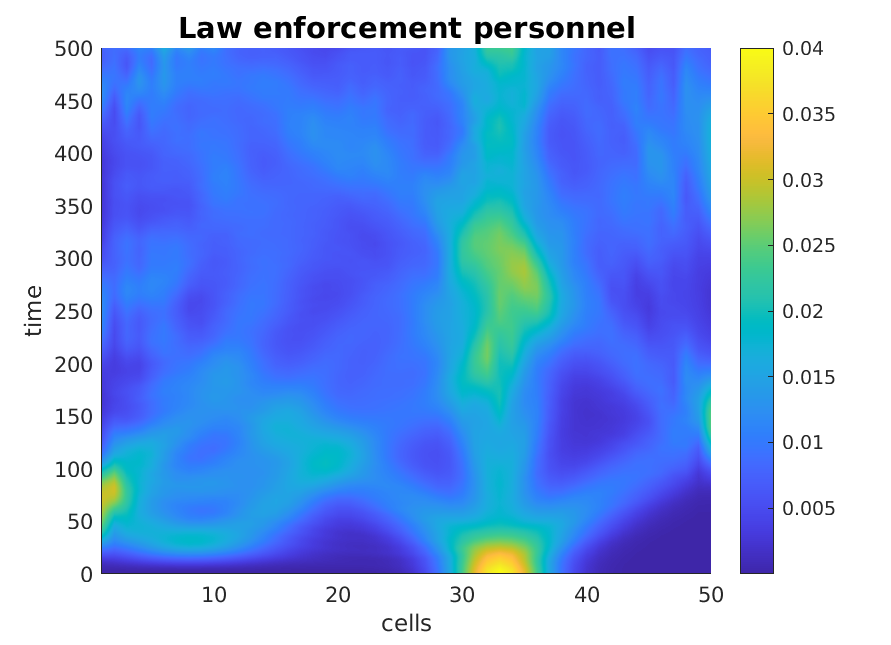}\label{fig:ic-3-om2-druganorule-c}}\\
\subfigure[]{\includegraphics[width=0.32\textwidth]{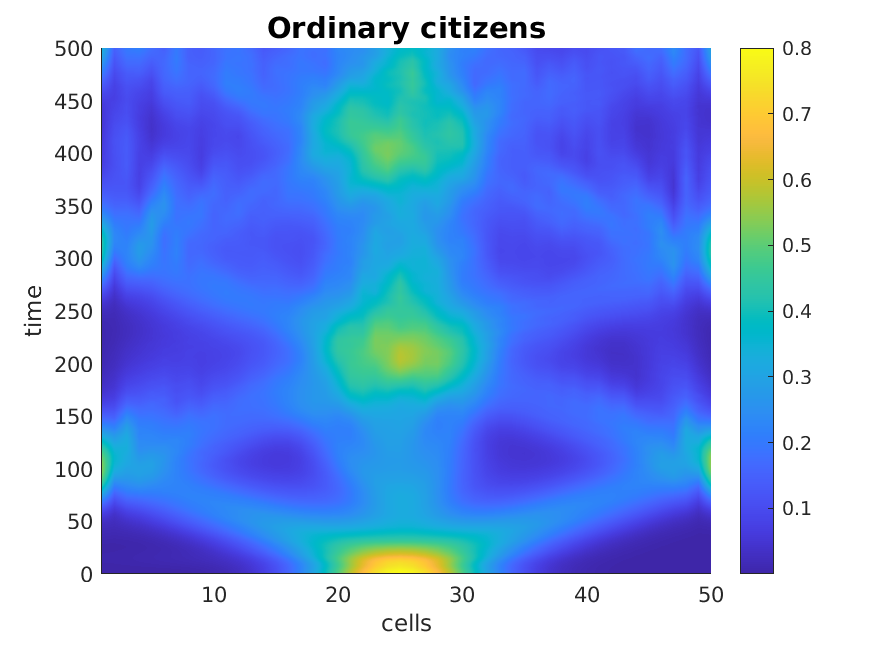}\label{fig:ic-3-om2-druganorule-d}}
\subfigure[]{\includegraphics[width=0.32\textwidth]{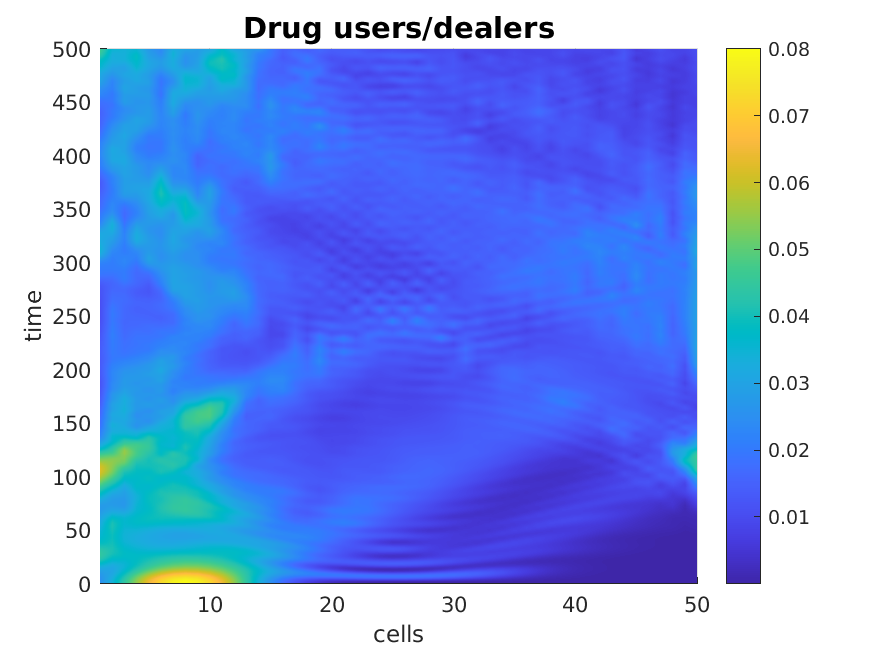}\label{fig:ic-3-om2-druganorule-e}}
\subfigure[]{\includegraphics[width=0.32\textwidth]{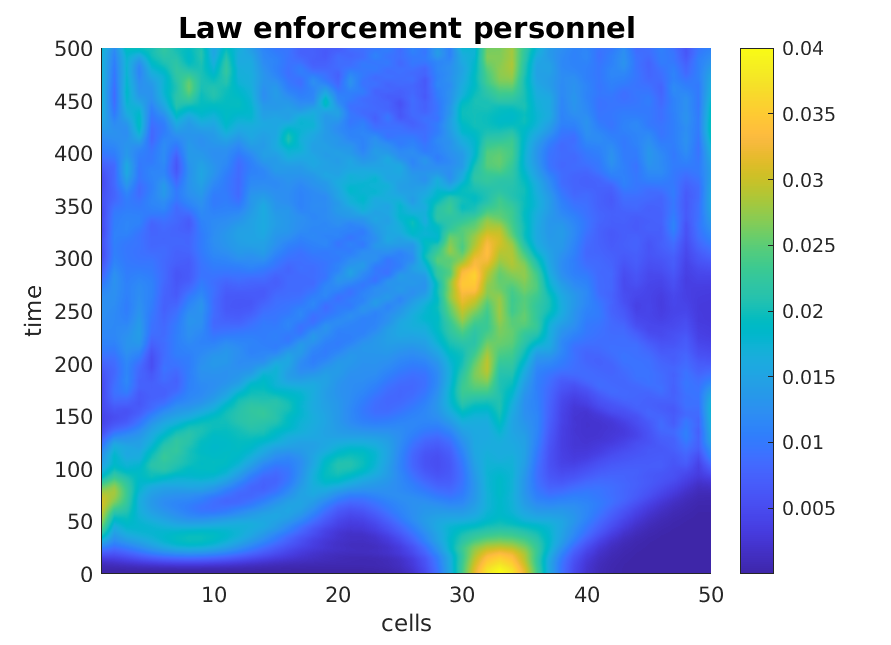}\label{fig:ic-3-om2-druganorule-f}}\\
\subfigure[]{\includegraphics[width=0.32\textwidth]{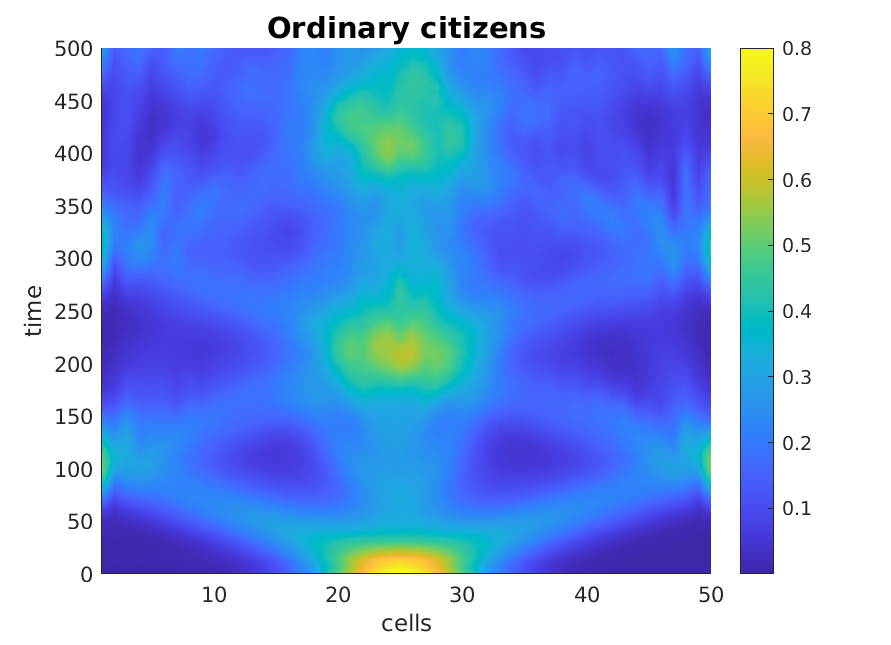}\label{fig:ic-3-om2-druganorule-g}}
\subfigure[]{\includegraphics[width=0.32\textwidth]{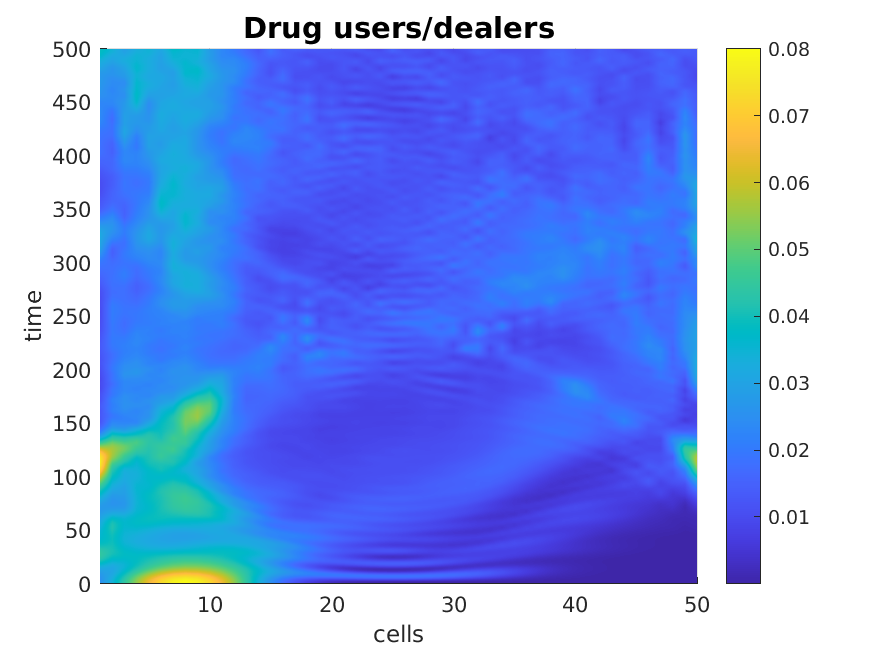}\label{fig:ic-3-om2-druganorule-h}}
\subfigure[]{\includegraphics[width=0.32\textwidth]{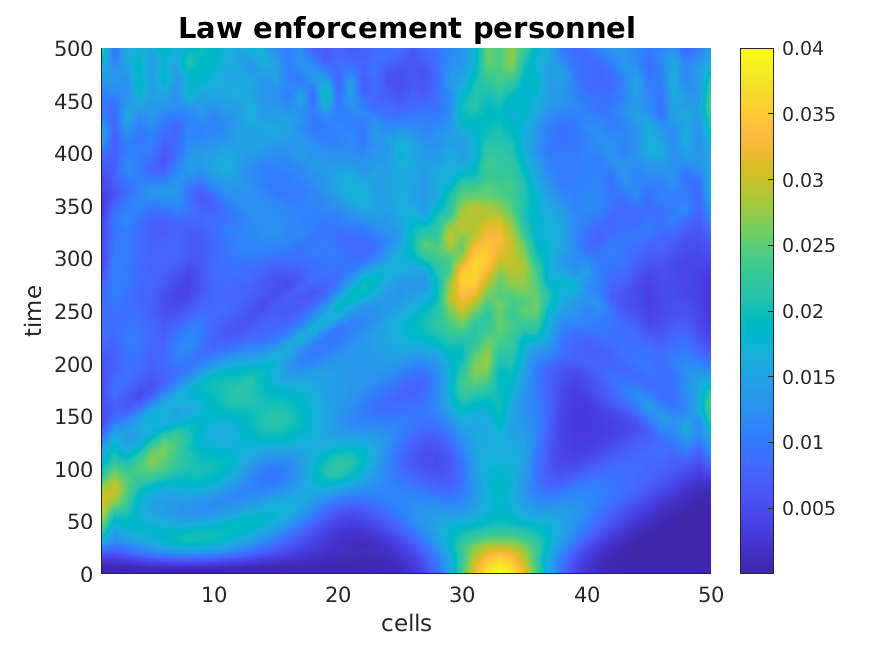}\label{fig:ic-3-om2-druganorule-i}}
\caption{Second scenario: time evolution and spatial distribution of the three subgroup densities. Ordinary citizens (\ref{fig:ic-3-om2-druganorule-a}, \ref{fig:ic-3-om2-druganorule-d} and \ref{fig:ic-3-om2-druganorule-g}), drug users/dealers (\ref{fig:ic-3-om2-druganorule-b}, \ref{fig:ic-3-om2-druganorule-e} and \ref{fig:ic-3-om2-druganorule-h}), and law enforcement personnel (\ref{fig:ic-3-om2-druganorule-c}, \ref{fig:ic-3-om2-druganorule-f} and \ref{fig:ic-3-om2-druganorule-i}). No-rule (upper), rule \eqref{rule_inertia} (center) and rules \eqref{rule_inertia}--\eqref{rule_diffusion} (bottom). The values of initial mean densities and parameters are given in \eqref{druginit3}, \eqref{Inertia2} and \eqref{drugpar1}.}\label{fig:ic-3-om2-druganorule}
\end{figure}

Some comments about the numerical simulations are in order.
Figures~\ref{fig:ic-2-om2-druganorule} and \ref{fig:ic-3-om2-druganorule}, where first and second scenario, \eqref{druginit2} and \eqref{druginit3}, respectively, are considered, show the time evolution of the mean densities of the three spatially distributed populations, considering both the standard Heisenberg view (Subfigures $(a),(b),(c)$) and the $(\mathcal{H},\rho)$--induced dynamics approach, where the rules \eqref{rule_inertia} and \eqref{rule_diffusion} (Subfigures $(d),(e),(f)$ and $(g),(h),(i)$, respectively) are used.

In both scenarios, when classical Heisenberg dynamics is used, we observe an oscillatory trend for all subgroups, as expected, since no rule is enforced. The results show similarities with Epstein's model \cite{epstein1997nonlinear}, despite the presence of oscillations. In \cite{epstein1997nonlinear}, drug users/dealers tend to move toward areas with a higher concentration of ordinary citizens, as these locations offer greater opportunities for committing crimes; law enforcement, in turn, intervenes in response to the presence of drug users/dealers, attempting to reduce their numbers and restore order.
This continuous cycle of interactions among criminals, citizens, and law enforcement can generate oscillations in both crime distribution and police intervention areas, similar to the oscillatory behavior observed in Figures~\ref{fig:ic-2-om2-druganorule} and \ref{fig:ic-3-om2-druganorule}. Even in the absence of explicit rules, the feedback among these agents-criminals moving toward citizens, citizens distancing themselves from drug users/dealers, and police reacting can explain the observed dynamics.
The phenomenon can be interpreted as a cyclic sequence of phases where crime increases, peaks, and is subsequently countered by police intervention, forming a self-sustaining, dynamically evolving system.
It is worth of being remarked that, since the initial inertia parameters have been choosen proportional to the peaks of the initial mean densities of the three subgroups, the tendency of ordinary citizens to remain in their affluent neighborhood is higher with respect to the case of using fixed values for the parameters across the entire domain, as commonly found in the literature \cite{inferrera2022operatorial,
gargano2021population,gorgone2023fermionic}. This leads to a significant difference when the classical Heisenberg dynamics is applied: since the inertia parameters assume higher values in areas of higher population densities, the three subgroups show a stronger tendency to remain in their current state at the density peaks. This results in greater resistance to change, leading to the emergence of small stationary pattern formations, although for very short periods of time.

By means of the $(\mathcal{H},\rho)$--induced dynamics approach, the standard Heisenberg dynamics is modified through the rules introduced in Section~\ref{sec:H-rho}, and the time evolution of the three subgroups accordingly changes (see Figures~\ref{fig:ic-2-om2-druganorule} and \ref{fig:ic-3-om2-druganorule}). By using the rule~\eqref{rule_inertia} (see Subfigures $(d),(e),(f)$), involving only modifications of the inertia parameters, the system exhibits a dynamics where oscillations are more controlled, and some patterns remain stable for short periods of time; this behaviour was not  possible before. This trend represents a significant first step toward approaching the patterns formation found in \cite{inferrera2024reaction}.
Thus, the obtained outcome can be interpreted as a formation of spatially stationary patterns but dynamically evolving over time, where the population distribution appears stably structured in space but subject to variations and oscillations. This dynamic complexity could represent an intermediate step allowing the formation of more sophisticated and coherent patterns, similar to Turing ones \cite{Turing}, which balance spatial instabilities with time fluctuations.
In Subfigures $(g),(h),(i)$ (see Figures~\ref{fig:ic-2-om2-druganorule} and \ref{fig:ic-3-om2-druganorule}), where rules \eqref{rule_inertia} and \eqref{rule_diffusion} modify both inertia and diffusion parameters, we note that
a trend similar to the case where only the rule \eqref{rule_inertia}, acting on inertia parameters, occurs. Anyway, the key difference is that, in this situation, law enforcement personnel are not only attracted towards areas with high criminals concentration but also to areas where ordinary citizens density increases. This because  law enforcement personnel can be attracted not only toward zones with high criminal activity but also to densely populated areas, for instance in situations involving public events, such as demonstrations, or strikes.
In other words, their actions are no longer exclusively focused on crime prevention and suppression but also on managing public safety in contexts where a high density of people may occur. This broader and multifaceted approach may reflect a strategy of law enforcement personnel aimed to maintain order and security in high-visibility situations susceptible of rising social risk, thus contributing to a more complex and varied dynamics compared to the model based solely on criminal influence. Additionally, we observe the emergence of stripe-like structures, although some small oscillations persist, which cannot be entirely suppressed as they are intrinsic to the construction of the operatorial model through the quantum-like approach.

\section{Conclusions}\label{sec: conclusions}
In this paper, we implemented an operatorial version of a model for crime phenomena, originally introduced by Epstein \cite{epstein1997nonlinear} and further investigated in \cite{inferrera2024reaction}. The population is partitioned in three subgroups that are modeled by fermionic ladder operators whose time evolution is assumed to be ruled by a Hermitian time-independent quadratic Hamiltonian  operator. The three subgroups exhibit either a local or a nonlocal competition, and are able to spread in a one--dimensional spatial region. The mean values of the number operators corresponding to the three subgroups in a cell can be interpreted as a measure of their local density in the cell. In order to have a more realistic dynamics, we used the $(\mathcal{H},\rho)$--induced dynamics approach, and suitable choices for the rules are given and motivated. The rules can modify the inertia and diffusion parameters, according to the evolution of the system. Two scenarios, characterized by different initial spatial distributions of the three subgroups, are considered, and some numerical results, together with their social interpretation, have been presented. Several extensions and generalizations of this model can be considered: effects due to cooperative interactions among the three subgroups can be incorporated in the Hamiltonian operator \cite{gorgone2023fermionic}, as well as a spatial model in a two--dimensional setting, with the aim of investigating if some stable patterns may emerge, can be considered. Work is in progress in these directions.

\clearpage

\section*{Acknowledgements}
This work has been carried out with the patronage of ``Gruppo Nazionale per la Fisica Matematica'' (GNFM) of ``Istituto Nazionale di Alta Matematica'' (INdAM). M.G.~acknowledges financial support from Finanziamento del Programma Operativo Nazionale (PON) ``Ricerca e Innovazione'' 2014-2020 a valere sull'Asse IV ``Istruzione e ricerca per il recupero''- Azione IV - Dottorati e contratti di ricerca su tematiche dell'innovazione, CUP J45F21001750007. M.G~acknowledges financial support from PRIN 2022 ``Nonlinear phenomena in low dimensional structures: models, simulations and theoretical aspects'', grant n. 2022TMW2PY. G.I.~acknowledges financial support from the project ``Reclutamento di Ricercatori di tipo A ed Assegnisti di Ricerca, Universit\`a degli Studi di Messina (Regione Sicilia)'', CUP
G41I22000690001. C.F.M.~acknowledges financial support from ``National Centre for HPC, Big Data and Quantum Computing (HPC)'', Codice progetto CN00000013 -- SPOKE 10, CUP D43C22001240001.

\end{document}